\documentstyle[12pt]{article}

\newcommand{\be}{\begin{equation}}
\newcommand{\ee}{\end{equation}}
\newcommand{\bea}{\begin{eqnarray}}
\newcommand{\eea}{\end{eqnarray}}
\newcommand{\nn}{\nonumber}

\newcommand{\p}[1]{(\ref{1})}

\begin{document}
\hspace*{3cm}
\begin{center}{\Large\bf
 Constrained Quantization on Symplectic Manifolds and Quantum Distribution
Functions}\\
\vspace*{10mm}
{\bf G.\ Jorjadze }\\
{\it Dept. of Theoretical Physics\\
Tbilisi Mathematical Institute\\
M.Aleksidze 1, 380093, Tbilisi\\
GEORGIA}\\
e-mail: jorj@ictp.trieste.it \\
and      jorj@imath.acnet.ge\\
\vspace*{1cm}
\end{center}

\begin{abstract}
{A quantization scheme based on the extension of phase space
with application of constrained quantization technic is considered.
The obtained method is similar to the geometric quantization.
For constrained systems the problem of scalar product on the
reduced Hilbert space is investigated and possible solution of this
problem is done. 
Generalization of the Gupta-Bleuler like conditions 
is done by the minimization of quadratic fluctuations
of quantum constraints. The scheme for the construction of
generalized coherent states is considered
and relation with Berezin quantization is found.
The quantum distribution functions are introduced and 
their physical interpretation is discussed.} \end{abstract}

\newpage
{\large \bf Introduction}

\vspace*{5mm}
\noindent
It is well known that the standard canonical quantization is not the universal
method for the quantization of Hamiltonian systems. Actually this method
is applicable only for the systems with a phase space having the cotangent
bundle structure.  For the generalization of canonical quantization 
different methods were developed and the geometric quantization [1] is
accepted as the most general one.

In [2] it was proposed a quantization scheme based on the extension
of phase space with further application of constrained quantization method
[3].
The obtained quantization turned out very similar to the geometric one.
The present work is the continuation of the activity started in [2].

The similar method with extension of phase space was introduced in [4a],
where for the quantization of constrained extended system the BFV (BRST)
quantization was used. In [4a] one can find also a wide variety of references
to different quantization methods and their short  analyses. Among 
other recent 
papers, which also use some extension procedure, it should be noted
[4b] and [4c].

The present paper is organized as follows:

In Section 1 the extended system is introduced. The phase 
space of the extended system is a cotangent bundle over the initial symplectic
manifold ${\cal M}$. For the quantization of the extended system the canonical
method is used and the pre-quantization operators arise as the result of 
some natural operator ordering.

In Section 2 for the extended system the certain constrained surface 
$\Phi_{f_k}=0$  is introduced. The constraint functions 
$\Phi_{f_k}$   characterized by  some complete set of observables
$f_k~(k=1,...2N)$ of the initial system and they form the set of the second
class constraints. Further, the constraint operators 
$\hat\Phi_{f}$  
are introduced and for the restriction of the extended quantum system 
the Dirac's 
($\hat\Phi_{f}|\Psi\rangle=0$) and the Gupta-Bleuler like 
($(\hat\Phi_{f}+i\epsilon\hat\Phi_{g})|\Psi\rangle=0$) conditions are used.
Certainly, the Dirac's condition are used only for the half of commuting
to each other constraints and the same number of complex conditions is
used in Gupta-Bleuler case too. Here, the standard problems of constraint 
quantization arise and in Appendixes A and B the possible solution of these
problems is considered. In particular, in Appendix A the
scalar product problem of physical states is investigated.
For the solution of this problem the limiting procedure 
($\epsilon \rightarrow 0$) with normalized physical states is used.

In Section 3, illustrating the quantization scheme
described above, we consider two examples.   
The first one is a quantization on the plane and the second one on the
cylinder.

In Section 4 we generalize the Gupta-Bleuler like conditions.
For this we use the minimization of quadratic fluctuations of quantum 
constraints.
Technical part of this method is described in Appendix C. 
The obtained condition 
contains the constraint operators in second order, 
and for the physical wave functions they are elliptic type equation 
on the phase space.

In Section 5 we introduce the general coherent states, which are
related with some complete set of observables.
The coherent states are constructed as the functions on the phase space and,
at the same time, they are parameterized by the points of the phase space.
The coherent states form the over complete set of states and have some 
interesting properties. In particular, they
minimize uncertainties of observables just
they are related to.
At the end of the section we construct special coherent states on the cylinder
and study their behavior in the limit when the squeezing parameter 
tends to zero. In this limit we get the eigenstates of the angular-momentum.

In Section 6 we introduce the quantum distribution functions as the square
of the modulus of physical wave functions. We get some smooth distributions
on the phase space and these functions 
satisfy some elliptic type equation.  This equation specifies
the distribution functions for the pure states. The generalization 
for mixed states is done as the convex
combination of pure ones. 
There are different classes of quantum distributions functions 
and each class is 
related to a certain complete set of observables of the system in 
consideration.
We discuss the physical interpretation of
the introduced distribution functions.
Namely, we interpret them as  the distributions
obtained in the experiment with simultaneous measurement
of observables which define the given class. 
At the end of the paper we discuss the
possibility for the formulation of quantum mechanics in terms
of quantum distribution functions without referring to the Hilbert space
and the operator formalism.

\vspace*{1cm}

\section{Quantization on a Cotangent Bundle}
\setcounter{equation}{0}

\noindent
We start with an introduction of some standard notations
of the Hamiltonian dynamics (see for example [1]).

The phase space of a classical system is a symplectic manifold 
${\cal M}$ and $\xi^k$, $(k=1,...,2N)$ are some local 
coordinates on  ${\cal M}$. For simplicity, the symplectic form 
$\omega = 1/2~\omega_{kl}(\xi )d\xi^k\wedge
d\xi^l $ is assumed to be exact: $\omega = d\theta$. 
Thus, $\omega_{kl} =\partial_k\theta_l-\partial_l\theta_k$, where
$\theta_k(\xi )$ are components of a 1-form
$\theta=\theta_k(\xi ) d\xi^k$.

Observables are smooth real functions on ${\cal M}$, and
the set of all observables ${\cal O}({\cal M})$ has the natural
Poisson-Lie structure.

The Hamiltonian vector field constructed for an observable $f(\xi )$
is given by
\be
V_f ={\cal V}_f^k \partial_k,~~~~ {\mbox {with}}~~~~
{\cal V}_f^k = \omega^{kl}\partial_l f
\ee
where $\omega^{kl}$ is the inverse (to $\omega_{kl}$) matrix:
$\omega^{ij}\omega_{jk} = \delta^i_k$. This field generates one-parameter
family of local canonical transformations.  

The Poisson bracket of two observables $f$ and $g$ is defined by
\be
\{ f, g\} \equiv  2\omega(V_f,V_g) =
- \partial_k f \omega^{kl} \partial_l g
\ee
and for global coordinates we have
$$
\{ \xi^k, \xi^l\} = - \omega^{kl}(\xi )
\eqno {(1.2')}
$$

The Hamilton function  $H=H(\xi )$ generates the dynamics 
of a system through the Hamilton's equations
$$
\dot {\xi}^i = {\cal V}_H^i (\xi )
$$
and this equations can be obtained by  variation of the action
\be
S = \int {[\theta_k (\xi )\dot {\xi}^k - H(\xi )]}dt
\ee

If the Hamiltonian system is constructed from the non-singular
Lagrangian [3], then the phase space ${\cal M}$ is a cotangent 
bundle over the configuration space of the corresponding Lagrangian system. 
In that case we have a separation of all coordinates $\xi^k, (k=1,...,2N)$ 
into two canonically conjugated parts. The first part is formed by
``coordinates" ($q^\alpha$) of the configuration space
and the second by ``momenta" ($p_\alpha$) ($\alpha =1,...,N)$.
The latter are unbounded ($-\infty <p_\alpha < +\infty $) and we can
use the standard scheme of canonical quantization with the rule:
\be
p_\alpha \longrightarrow {\hat p}_\alpha = -i\hbar\frac{\partial}
{\partial q_\alpha}
\ee

According to Darboux's theorem, the canonical coordinates exist
on an arbitrary symplectic manifold; but in general, such coordinates
exist only locally [1], and there is no global cotangent bundle
structure with unbounded momenta. Consistent quantization requires
a realization of not only the classical commutation relations, but of
spectral conditions as well. Respectively, in general,
the rule (1.4) is not acceptable, since the spectra of the 
differential operators are unbounded. 

Note, that a symplectic manifold of general type naturally
arises for the systems with singular Lagrangian
(for example for gauge theories), when we apply the Dirac's procedure
for constrained dynamics [3].

To generalize quantization method for such cases too 
we introduce some auxiliary
Hamiltonian system with the phase space $T^*{\cal M}$, where
$T^*{\cal M}$ is the cotangent bundle over the symplectic manifold
${\cal M}$. The new system has $4N$ dimension,
and we choose the 1-form  $\Theta=P_kd\xi^k$,
where $(P_k, \xi^k)$ are the standard coordinates on 
the cotangent bundle $T^*{\cal M}$: $ P_k = P(\partial_{\xi^k})$.
So, the coordinates $P_k$ play the role of ``momenta", while the
$\xi^k$ are ``coordinates". The corresponding symplectic form 
is canonical:
$d\Theta = dP_k\wedge d\xi^k$, and for the Poisson
brackets of the new system we have (compare with (1.2$'$))
\be
\{ \xi^k, \xi^l\}_* = 0 = \{ P_k, P_l\}_* 
~~~~~\{ P_k, \xi^l\}_* = \delta_k^l 
\ee
The index $*$ is used to make difference between the 
Poisson brackets (1.2) and (1.5).
Below we denote the initial system by $ M$, 
and the extended new system by $T^* M$.
  
Since the symplectic form $\omega$ is non-degenerated, the relation
\footnote {$ \omega (\Phi ,~\cdot ~) $ denotes the contraction of
$\omega $ with  $\Phi $: $\omega (\Phi ,~\cdot ~)_l= \Phi^k\omega_{kl}$.}
\be
  \omega (\Phi ,~\cdot ~)= (\theta - P)~(\cdot )
\ee
defines the vector field $\Phi$ ($\Phi \in V({\cal M})$) uniquely.
The components of this field $\Phi^k$ are given by   
$$
  \Phi^k = \omega^{kl} (P_l - \theta_l)
\eqno (1.6')
$$
and respectively, we get the map  ($T^*{\cal M} \mapsto V({\cal M})$)  
of the cotangent bundle $T^*{\cal M}$ to the  space of 
vector fields on ${\cal M}$. 
Using this vector field $\Phi $ and some observable 
$f(\xi ) \in {\cal O}({\cal M})$ we can construct
the function
$\Phi_f$ on $T^*{\cal M}$
\be
\Phi_f \equiv \Phi(f) = \Phi^k\partial_k f
\ee
and from (1.6$'$) we have
$$
\Phi_f = \theta (V_f) -P(V_f)
\eqno (1.7')
$$
where $V_f$ is the Hamiltonian vector field (1.1).

The definition of functions $\Phi_f $ by (1.7) at the same time
gives the map 
$$ {\cal O}({\cal M}) \mapsto
{\cal O}({T^*\cal M})$$ 
of observables of the system 
$ M$ to the certain class of functions  on 
$T^*{\cal M}$. Then, from (1.5 -1.7$'$) we obtain
\be
\{ \Phi_f,\Phi_g\}_* 
=-\{ f, g\} - \Phi_{ \{ f, g\} }~~~~
\{\Phi_f, g\}_* = -\{ f, g\} 
\ee
Note, that these commutation relations are written for the system $T^* M$,
and here for the functions $\{ f, g\}$ and $ g$
we use the same notations as for the corresponding observables on ${\cal M}$.
Strictly, of course, we should  distinguish between these functions.
However, it is generally simpler not to do this except in case of possible
confusion. 

Now, let us introduce a new map from 
${\cal O}({\cal M})$ to ${\cal O}(T^*{\cal M})$
\be
f \mapsto R_f \equiv f - \Phi_f
\ee
which in local coordinates $(P_k, \xi^k)$ takes the form
$$
R_f = f(\xi ) +\partial_kf(\xi ) \omega^{kl}(\xi ) (P_l - \theta_l(\xi ))
\eqno (1.9')
$$

The 1-form $\theta$ in (1.9$'$) is assumed to be fixed, and 
the map (1.9) defines the class
of observables $R_f$ uniquely. We have that $R_f\neq R_g$ whenever $f\neq g$.
Note that a change of the 1-form $\theta$ by an exact form $dF$:
$\theta_k(\xi ) \longrightarrow \theta_k(\xi ) + \partial_kF(\xi )$,
corresponds to
$$
R_f\rightarrow R_f + \{f,F\}
\eqno(1.9'')
$$
and for the system $T^*M$ it is the canonical transformation generated
by the function $F(\xi )$.

Then, using (1.8), for the Poisson brackets of constructed observables (1.9),
we obtain 
\be
\{ R_f, R_g\}_*  = R_{\{ f, g\} } 
\ee

We choose the Hamiltonian of the extended system $T^* M$ to be
equal to $R_H$, where $H=H(\xi )$ is the initial Hamiltonian.
Respectively, for the system $T^* M$ the action (1.3) 
takes the form
\be
S_{T^*{\cal M}} = \int {[P_k (\xi )\dot {\xi}^k - R_H (P,\xi )]}dt
\ee

The linear map (1.9) has two remarkable properties:\\ 
1. It preserves the Poisson brackets (see (1.10)).\\
2. The functions $R_f$ in (1.9) contain the momentum variables $P_k$
no higher than in the first degree.\\
Below we use these properties
 for the construction of the corresponding quantum operators.

As it was mentioned above, the system $T^* M$ can be 
quantized by the scheme of canonical quantization.
This means that the Hilbert space ${\tilde {\cal H}}$ is the space of squad
integrable functions $\Psi (\xi )$ on ${\cal M}$:
$\tilde {{\cal H}} = {\cal L}_2 ({\cal M})$.
It is convenient to introduce the invariant measure on ${\cal M}$
\be
d\mu (\xi )\equiv \sqrt {\omega(\xi )}~d^{2N}\xi
~~~~~{\mbox {with}}~~~~ \omega (\xi )\equiv det~\omega_{kl}(\xi )
\ee
and to define the scalar product by
\be
\langle\Psi_2 |\Psi_1 \rangle = \int d\mu (\xi )~ 
\Psi_2^* (\xi ) \Psi_1 (\xi )
\ee
According to the scheme of canonical quantization 
for the function $f(\xi )$ we have the corresponding  
operator $\hat f$ which acts  
on a wave function $\Psi (\xi )$ as the multiplication by $f(\xi )$.
Taking into account the remarks after the equations (1.8),
we use the same notation $ f(\xi )$ 
for this operator $\hat f$ as well: $\hat f \equiv f(\xi )$. 

Further, the rule (1.4) defines  the Hermitian operators
$\hat P_k$ 
\be
  {\hat P}_k = -i\hbar\partial_k
-i\hbar\frac{\partial_k\omega (\xi )}{4\omega (\xi )}
\ee
where the additional term, proportional to $\partial_k\omega$, 
arises from the measure (1.12) in (1.13).

Construction of Hermitian operators,
in general, has an ambiguity connected to the
ordering of coordinate and momentum operators 
in the functions of corresponding observables. 
For the functions $R_f$ this ordering
problem is only for the term 
$\partial_kf\omega^{kl}P_l$ (see (1.9$'$)).
When the momentum operator is only in the first 
degree, it is easy to see, that the following symmetric ordering
\be
\partial_kf\omega^{kl}P_l \longrightarrow
\frac{1}{2}
(\partial_kf\omega^{kl}{\hat P_l} +
{\hat P_l}\partial_kf\omega^{kl})
\ee
defines a Hermitian operator, and for those operators there
are no anomalies in the quantum commutation
relations. Now, choosing the ordering (1.15) in (1.9$'$), 
and using that \footnote {The formula (1.16) is a consequence of 
the Jacobi identity:
$\omega^{il}\partial_l\omega^{jk}+\omega^{jl}\partial_l
\omega^{ki}+ \omega^{kl}\partial_l\omega^{ij} = 0$}
\be
\partial_k(\sqrt {\omega}~\omega^{kl})=0
\ee
we obtain
\be
   {\hat R}_f = f(\xi ) - \theta (V_f) -i\hbar V_f 
\ee
where $V_f$ is the Hamiltonian vector field (1.1), and
$\theta (V_f)$
is the value of the 1-form $\theta $ on this field:
$\theta (V_f) = \theta_k\omega^{kl}\partial_lf$.
So, the operator $\hat R_f$ is constructed from the invariant terms,
and it does not depend on the choice of coordinates $\xi^k$ on $\cal M$.

Note, that a change of a 1-form $\theta$ by an exact form $dF$
corresponds to the unitary transformation of operators
$\hat R_f$ (see (1.9$''$)) 
$$
{\hat R}_f \longrightarrow 
e^{-\frac{i}{\hbar}F(\xi )}
{\hat R}_ f
e^{\frac{i}{\hbar}F(\xi )}
$$ 

Since the operator ordering (1.15)
avoids anomalies in the commutation relations, from (1.10) we get
\be
[{\hat R}_f,{\hat R}_g] =-i\hbar{\hat R}_{\{ f, g\} } 
\ee
and this is the most interesting point of the described quantization scheme
on the cotangent bundle of a symplectic manifold.

It is remarkable, that the operators (1.17) (which 
arise naturally in our scheme) are the 
pre-quantization operators of the theory of
geometric quantization, and a representation of Poisson brackets algebra
by these operators is a well known fact from this theory [1].

After canonical quantization on the cotangent bundle $T^*{\cal M}$
our goal is to use this quantum theory for the quantization of the 
initial system $M$,
and in the next section we consider the connection between these two systems.

\vspace*{1cm}

\section{Constraints on a Cotangent Bundle}
\setcounter{equation}{0}

\noindent
Geometrically there is a standard
projection  ($\pi: T^*\cal M \rightarrow \cal M $)
of the cotangent bundle $T^*\cal M$ to 
the initial phase space ${\cal M}$.
To find the dynamical relation between these two systems
we introduce the constraint surface on the cotangent bundle 
$T^*\cal M$, and define it as the kernel of 
the mapping $T^*{\cal M} \rightarrow V({\cal M})$ given by (1.6)-(1.6$'$).  
It means that on the constraint surface the vector field $\Phi $
vanishes: $\Phi  = 0$, and if we use the functions $\Phi_f (P,\xi )$
(see (1.7)) this surface can be written as
\be
\Phi_f~= 0 , ~~~~~ \forall~~ f(\xi )\in {\cal O}({\cal M}) 
\ee

From (1.8) and (1.9) we have
\be 
\{ R_f, \Phi_ g\}_*  = \Phi_{\{ f, g\} } 
\ee
and we see, that (2.1), i.e. the constraint surface, is
invariant under the canonical transformations generated by the functions
$R_f$. In particular it is invariant in dynamics
generated by the Hamiltonian $R_H$. Note, that the 1-form $\theta$ is assumed
to be fixed in all these formulas.

In local coordinates the surface (2.1) can be written as
\be
P_k - \theta_k (\xi ) = 0
\ee
(see (1.6$'$) -(1.7$'$)), and respectively, the momenta $P_k$ are 
defined uniquely. 
Hence, the coordinates $\xi^k (k=1,...2N)$ can be used for the
parameterization of the constraint surface, and this surface
is diffeomorphic to the manifold ${\cal M}$. Then, the reduction
procedure gives (see (2.3) and (1.9$'$))
$$
P_kd\xi^k|_{\Phi =0} =\theta_k(\xi )d\xi^k~~~~~~~
R_H|_{\Phi =0} = H(\xi )
$$
and the action (1.11) of the system  $T^* M$
is reduced to (1.3).
Thus, we conclude that the classical system $T^* M$ on the
constraint surface $\Phi_f = 0$ is equivalent to the initial one.

To find the connection on the quantum level too, we have
to quantize the system $T^* M$ taking into account
the constraints (2.1).

Before beginning the quantum part of the reduction scheme, let us note, that
the constraints (2.2) are written for an arbitrary observable $f(\xi )$,
and since the constraint surface $\Phi =0$ is $2N$ dimensional,
only the finite number of those constraints are independent.

To select the independent  constraints we introduce the complete
set of observables on $\cal M$.
The set of observables $\{ f_n(\xi )\in {\cal O} ({\cal M});
(n=1,...,K)\}$ is called complete, if any observable 
$f(\xi )\in {\cal O}({\cal M})$ 
can be expressed as a function of this set 
\be
f={\cal F} (f_1,...,f_K)
\ee
It is clear that $K\geq 2N$, and
we can choose the set with $K=2N$
only for the manifolds with global coordinates. 
For $K>2N$ there are some functional
relations for the set $f_1,...,f_K$, and locally
only 2N  of these functions are independent. 
Then, from (1.7) and (2.4) we have 
\be
   \Phi_f = \frac{\partial {\cal F} }{\partial f_n}\Phi_{f_n}
\ee
and the constraints (2.1) for arbitrary $f$ are equivalent to $K$ 
constraints
\be
\Phi_{f_n} =0,~~~~ (n = 1,...,K)
\ee
In particular, in case of global coordinates 
we can introduce only $2N$ constraints $\Phi_{f_n}, (n=1,...,2N)$.
If it is not specified, below we are assuming that a
manifold $\cal M$ has global coordinates and a set of functions
$f_1,...,f_{2N}$ is  complete. Note, that the constraint surface
and the reduced classical system are independent on the choice of
such complete set.
Using (1.8), we see that on the constraint surface (2.1)
the rank of the matrix $\{ \Phi_{f_n},\Phi_{f_m}\}_* $
is equal to $2N$, and therefore, these constraints, in Dirac's 
classification, are the second class constraints.

For the constrained systems there are, actually, two schemes
of quantization:\\
A. ``First reduce and then quantize".\\
B. ``First quantize and then reduce".

By the scheme A we are returning to the initial problem
of quantization on the manifold ${\cal M}$.
Therefore, it is natural to use the scheme B, especially as, the first
step of this scheme we have already accomplished.

To justify our strategy it is necessary to show,
that the schemes A and B  give equivalent quantum 
theories, when the system $ M$ is quantizable by the canonical
method, and also, it is worthwhile to have a certain general 
receipt for accounting the constraints (2.6)
on the quantum level. 

According to the scheme B the next step is a construction of 
constraint operators. 
From (1.9) and (1.17) the operators 
\be
\hat \Phi_f = i\hbar V_f + \theta (V_f) 
\ee
are Hermitian, and by direct calculation one obtains
$$
[{\hat \Phi}_f,{\hat \Phi}_g] =i\hbar (\{f,g\} +
 {\hat \Phi }_{\{ f, g\} }) 
\eqno (2.7')
$$ $$
[{\hat R}_f,{\hat \Phi}_g] =-i\hbar {\hat \Phi}_{\{ f, g\} } 
\eqno (2.7'')
$$
These  commutators are quantum versions of the relations (1.8) and (2.2).
As it was expected, there are no anomalies for them (see (1.18)).

Now, we should make reduction of Hilbert space using the constraint
operators (2.7) for some complete set of functions $f_1,...,f_{2N}$.
The reduced Hilbert space for the constrained systems is called 
the physical Hilbert space as well, and we denote it by ${\cal H}_{ph}$.

For systems with the second class constraints there is the 
following reduction procedure  [3]:
one has to select a commuting subset of $N$ constraints 
\footnote {If subset of constraints is treated as the first
class (independently from others), then, in our case,
they are commuting (see (2.7$'$)).}
$\hat \Phi_1,...,\hat \Phi_N$
$$
 [\hat \Phi_a,\hat \Phi_b] = 0~~~~~~~~~1\leq a, b \leq N
$$
and then, construct a physical
Hilbert space ${\cal H}_{ph}$ from the states
which satisfy the Dirac's conditions
$\hat \Phi_a |\Psi_{ph}\rangle =0,~ a=1,...,N$.
Note, that we can not put all constraints equal to zero in strong sense
($\hat \Phi_k |\Psi\rangle =0,~ k=1,...,2N$), since it contradicts to 
commutation relations of the second class constraints.

From (2.7$'$) we see that in our case, the described procedure implies
selection of $N$ commuting observables $f_a,~ a=1,...,N; 
\{f_a, f_b\}=0$, and further, solution of the differential equations 
\be
\hat \Phi_{f_a}~ \Psi_{ph} (\xi ) =0,~~~~~~~ a=1,...,N
\ee
Construction of physical states by 
selection of $N$ commuting observables 
is quite natural from the point of view of standard quantum mechanics, 
and we shall return to this point later.

Equations (2.8) are the first order linear differential equations and, 
in principle, they can be explicitly integrated to describe corresponding
wave functions.
But at this stage of quantization scheme B
two significant problems usually arise:
the first is connected with the introduction of scalar product for the 
physical vectors [5],
and the second, with the definition of 
observable operators on these vectors.
 
For the first problem, the point is, that solutions of Dirac's conditions 
$\hat {\Phi}_a |\Psi_{ph}\rangle =0$, in general,
are not the vectors of the same Hilbert space where the first stage of
quantization was accomplished (in our case ${\cal L}_2({\cal M})$),
and it is necessary to introduce the structure of Hilbert space
additionally.
These solutions, as a rule, are in the space dual to the Hilbert space,
and one has to introduce the new scalar product for them.

In our case, solutions of (2.8), in general, are not square integrable
on $\cal M$ (usually they are generalized functions),
and the scalar product (1.13) needs modification.
 On the other hand, a certain measure in scalar product
defines the class of functions square integrable by this measure.
Thus, a measure for the new scalar product and the class of solutions
of (2.8) should be adjusted.

One method for the solution of this problem is based on the 
introduction of complex constraints [6].
Note, that classical observables $f(\xi )$ are assumed to be real functions 
on a phase space, but it is clear, that the whole considered
construction (except for the self-adjointness) can be naturally extended
for complex valued functions $f(\xi )=f_1(\xi )+if_2(\xi )$
as well.

Using the remaining part of constraints 
$\Phi_{f_{N+1}},...,\Phi_{f_{2N}}$,
one can introduce constraints for the complex functions
$Z_a = f_a + i\epsilon f_{N+a}$
and consider the equations
\be
(\hat {\Phi}_{f_a} + i\epsilon\hat {\Phi}_{f_{N+a}}) |\Psi_\epsilon\rangle =0
~~~~~~~~a=1,...,N
\ee
Here, $1\leq a\leq N,~ \{f_a, f_{N+a}\} \neq 0 $
and $\epsilon $ is some real parameter \footnote {Sometimes we omit
the index ``ph" for the physical vectors
(and physical Hilbert space), 
and use the index $\epsilon$ only}.

The condition (2.9) looks like Gupta-Bleuler quantization [7], and 
for normalizable solutions $|\Psi_\epsilon\rangle$ the 
mean values of corresponding constraints vanish
\be
\langle\Psi_\epsilon |\hat \Phi_{f_a}|\Psi_\epsilon\rangle =0~~~~~~~~~~~~~~
\langle\Psi_\epsilon |\hat \Phi_{f_{N+a}}|\Psi_\epsilon\rangle =0
\ee 

It turns out that the solutions of (2.9) could be square integrable indeed,
and then, they form some subspace of the Hilbert space ${\cal L}_2({\cal M})$
(see the below). 
The corresponding reduced physical Hilbert space we denote by 
${\cal H}_\epsilon$.
We have 
$\Psi_\epsilon (\xi )\in {\cal H}_\epsilon 
\subset {\cal L}_2({\cal M})\subset {\cal L}^*_2({\cal M})$, where 
${\cal L}^*_2({\cal M})$ is the space dual to the Hilbert
space ${\cal L}_2({\cal M})$. 
If we consider the physical states $|\Psi_\epsilon \rangle$ 
as the vectors of the dual space
${\cal L}^*_2({\cal M})$, then
the suitable choice of the norms $||\Psi_\epsilon ||$, and 
some smooth dependence on the parameter $\epsilon $
can provide existence of the limit
$$
\lim \limits_{\epsilon \to 0} |\Psi_\epsilon \rangle =
|\Psi_{ph} \rangle 
$$
where $|\Psi_{ph} \rangle\in {\cal H}_{ph} \subset{\cal L}^*_2({\cal M})$
(see Appendix A).
Obtained physical states $|\Psi_{ph}\rangle$  
specify the class of solutions of (2.8),
and the scalar product
for them is defined by
\be
\langle\Psi_{2ph} |\Psi_{1ph}\rangle =
\lim \limits_{\epsilon \to 0} \frac 
{\langle\Psi_{2\epsilon}|\Psi_{1\epsilon}\rangle }
{||\Psi_{2\epsilon}||~||\Psi_{1\epsilon}||}
\ee
where $|\Psi_{1ph}\rangle $ and $|\Psi_{2ph}\rangle $
are the limits of $|\Psi_{1\epsilon}\rangle$ and $|\Psi_{2\epsilon}\rangle$ 
respectively.
Note, that in the limit $\epsilon \rightarrow 0$
the norm of vectors $||\Psi_{\epsilon}||$ usually diverges, but the
scalar product (2.11) remains finite
($|\langle\Psi_{2ph} |\Psi_{1ph}\rangle| \leq 1$). 
(for more details see Appendix A and the examples in the next section).

It is remarkable that the choice of physical states by the conditions
(2.8) and (2.9) is equivalent
to the choice of real and complex polarizations 
of geometric quantization respectively [1].

The second above mentioned problem is connected with the fact,
that a reduced Hilbert space constructed by (2.8) (or (2.9)),
in general, is not invariant
under the action of some pre-quantization operator 
${\hat R_g}$. Indeed, the invariance conditions for (2.8) 
are 
\be
[\hat {R}_g ,\hat {\Phi}_{f_a}]=
 \sum_{b=1}^{N}~d^b_a \hat {\Phi}_{f_b}~~~~~~~(1\leq a\leq N)
\ee 
and, from (2.7$''$) we see that it is not valid for arbitrary $g(\xi )$.
Moreover, even if a pre-quantization operator 
acts invariantly on ${\cal H}_{ph}$, this operator can be non-Hermitian
on ${\cal H}_{ph}$, when the latter is not a subspace
of ${\cal L}_2({\cal M})$ and the Hilbert structure is introduced
additionally (see the example below).

For the definition of the corresponding observable operator on 
the physical Hilbert space one can deform the
pre-quantization operator adding quadratic (and higher)
powers of constraint operators \footnote {Corresponding procedure
in classical case is given in Appendix B}. 
Then, using commutation relations
(2.7$'$) and (2.7$''$), one can construct a new Hermitian operator,
which is invariant on the reduced Hilbert space.
Of course, there are different possible deformations, and in
general, they define different operators on the physical Hilbert space.
In terms of usual canonical quantization, different deformations
correspond to different operator orderings.
This is the standard ambiguity of quantum theories which 
in the classical limit $\hbar \rightarrow 0$ vanishes.
Note, that corresponding deformed classical functions
are indistinguishable on the constraint surface $\Phi_f =0$.

The described quantization scheme we call E-quantization scheme.
In the next section we consider application of this
scheme for some simple examples. 
We use these examples
as a test for our approach as well.

\vspace*{1cm}

\section{Examples of E-Quantization scheme}
\setcounter{equation}{0}

\noindent
Example 1. Phase space is a plane ${\cal M} \equiv {\cal R}^2$
with standard coordinates $\xi^1 \equiv p,~ \xi^2 \equiv q$ 
and the symplectic form $\omega = dp\wedge dq$.
The coordinates $p$ and $q$ are global
and from (1.7$'$) we get
\be
\Phi_p = \frac {1}{2}p -P_q 
~~~~~~~~~~\Phi_q = \frac {1}{2}q +P_p 
\ee
where, for the convenience, we choose the 1-form $\theta =\frac {1}{2}pdq
- \frac {1}{2}qdp$. 
The corresponding constraint operators are
\be
{\hat \Phi}_p = \frac {1}{2}p +i\hbar\partial_q 
~~~~~~~~~~{\hat \Phi}_q = \frac {1}{2}q-i\hbar\partial_p 
\ee
and, according to (2.9), for the physical vectors $|\Psi_\epsilon\rangle$ 
we have the equation
\be
\left ( {\hat \Phi}_q - i\epsilon {\hat \Phi}_p\right ) 
|\Psi_\epsilon\rangle =0
\ee
with some positive parameter $\epsilon $ ($\epsilon > 0$)
\footnote {For $\epsilon \leq 0$ equation (3.3) has no normalizable
solutions.}.
Solutions of (3.3) have the form
\be
 \Psi_\epsilon (p,q) = 
\exp{(-\frac{\epsilon p^2}{2\hbar})}
\exp{(-\frac{i pq}{2\hbar})}
\psi (q-i\epsilon p)
\ee
where $\psi $ is an arbitrary function.
For the square integrability of these solutions  
we can specify the class of $\psi $ functions, for example, by
\be
 \psi (\xi )= 
\exp{(-\frac{\gamma \xi^2}{2})}P(\xi )~~~~~~~(\xi \equiv q-i\epsilon p)
\ee
Here $\gamma $ is some fixed positive parameter ($\gamma > 0$), and
$P(\xi )$ --- any polynomial. 
Then, for sufficiently small $\epsilon$
the functions (3.4) will be square integrable on the plane and 
they form the physical subspace 
${\cal H}_\epsilon$, 
(${\cal H}_\epsilon \in {\cal L}_2({\cal R}^2)$). 

To investigate the case $\epsilon =0$, we consider the limit 
$\epsilon \rightarrow 0$ of functions (3.4) (see Appendix A), and get
\be
\Psi_{ph} (p,q) = 
\exp{(-\frac{i pq}{2\hbar})}~\psi (q)
\ee
It is clear, that these functions are not squared integrable on the plane,
but they are well defined elements of the dual space 
$\Psi_{ph}(p,q)\in{\cal L}^*_2({\cal R}^2)$. 
The functions (3.6) form the physical Hilbert space ${\cal H}_{ph}$,
and they are solutions of (3.3) with $\epsilon =0$.
Using  rule (2.11), we obtain
\be
\langle\Psi_{2ph}|\Psi_{1ph}\rangle = \frac{1}{N_1N_2}
\int \psi^*_2 (q)\psi_1(q)~ dq 
\ee
where 
$$
N^2_i =\int |\psi_i (q)|^2~ dq ~~~~~~ (i=1,2;~~ N_i >0) 
$$

Action of pre-quantization operators 
\be
{\hat R}_p = \frac {1}{2}p -i\hbar\partial_q 
~~~~~~~~~~{\hat R}_q = \frac {1}{2}q+i\hbar\partial_p 
\ee
on the physical states (3.6)  gives
\be
{\hat R}_p \Psi_\epsilon (p,q)=
\exp{(-\frac{i pq}{2\hbar})}~(-i\hbar) \psi '(q)~~~~~~~
{\hat R}_q \Psi_\epsilon (p,q) = 
\exp{(-\frac{i pq}{2\hbar})}~q\psi (q)
\ee
Thus, from (3.7) and (3.9) we have the standard 
coordinate representation of quantum mechanics. 
Similarly, one can obtain the momentum representation
in the limit $\epsilon \rightarrow \infty $ with
corresponding choice of the class of solutions (3.5).

Let us  consider the problem of construction of some observable operators on the 
physical Hilbert space 
${\cal H}_{ph}$ (3.6). 
It is easy to check that this space is invariant under the action of
pre-quantization operators $\hat R_f$,
where  $f(p,q)=pA(q)+U(q)$, with arbitrary $A(q)$ and $U(q)$.
But it turns out, that these operators $\hat R_f$
are Hermitian 
(with respect to the scalar product (3.7)) only for the constant 
function $A(q)$ ($A(q)=c$).
Similarly, there is a problem of definition of kinetic energy
operator, since the corresponding pre-quantization operator is not
defined on the chosen ${\cal H}_{ph}$ 
\footnote {For this ${\cal H}_{ph}$, such problem
have functions containing momentum $p$ in second 
and higher degrees}. 
These are just the problems mentioned at the end of the previous section, 
and for the definition of corresponding observable operators
we can make appropriate deformations (see Appendix B). 
For example, deformation
of the pre-quantization operator of kinetic energy $E=p^2/2m$
by the quadratic term
$$
 {\hat R}_{p^2/2m} \rightarrow 
 {\hat R}_{p^2/2m} + 
\frac {1}{2m}\hat \Phi_p^2 \equiv \hat E
$$
gives that the corresponding operator $\hat E$ is well defined 
on ${\cal H}_{ph}$, and effectively it acts as the standard kinetic 
energy operator 
$$
 {\hat E} : \psi (q) \mapsto -\frac {\hbar^2}{2m}\psi ''(q)
$$

Now, we return to the physical subspace ${\cal H}_\epsilon $ with some fixed
positive $\epsilon$. In complex coordinates
\be
z=\frac {q + i\epsilon p}{\sqrt {2\epsilon\hbar}}~~~~~~
z^*=\frac {q - i\epsilon p}{\sqrt {2\epsilon\hbar}}
\ee
(3.3) takes the form 
$$
\left (\partial_z + \frac {z^*}{2}\right ) \Psi_\epsilon (z,z^*) =0
\eqno (3.10')
$$
and the solutions are 
\be
\Psi_\epsilon (z,z^*) = \exp{(-\frac {1}{2}|z|^2)}F (z^*)
\ee 
where $F (z^*)$ is any holomorphic function of $z^*$.
Comparing (3.11) and (3.4) we have
$ F(z^*) =\exp {(1/2~{z^*}^2)}~ \psi (\sqrt {2\epsilon\hbar}~z^*)$.
From the point of view of canonical quantization
the complex coordinates $z$ and $z^*$ (see (3.10)) 
are the classical functions of annihilation and creation operators
$\hat {\bf a}$ and $ {\bf \hat a^*}$ respectively.
The corresponding pre-quantization operators
$$
 \hat R_z = \frac {z}{2} +\partial_z^* 
~~~~~~~~~~~~ \hat R_{z^*} = \frac {z^*}{2} -\partial_z
$$
act invariantly on the physical Hilbert space 
$\cal {H}_\epsilon$, and we have
$$
\hat {R}_z \Psi_\epsilon (z,z^*) =\exp{(-\frac {1}{2} |z|^2)}
F'(z^*)
~~~~~~~~~~~~
\hat R_{z*} \Psi_{ph} (z,z^*) = \exp{(-\frac{1}{2}|z|^2)}z^*F(z^*)
$$
Thus, the reduction on $\cal {H}_\epsilon$ gives the holomorphic 
representation of quantum mechanics [8], and we see that for the Example 1 
the quantum theory of the E-quantization scheme is equivalent to the 
ordinary canonical one.
For different $\epsilon $ the physical Hilbert spaces $\cal {H}_\epsilon$
are different subspaces of ${\cal L}_2({\cal R}^2)$, and the corresponding
representations of canonical commutation relations
are unitary equivalent due to Stone - von-Neumann theorem [9].

Further, for any state $|\Psi\rangle$ of standard quantum mechanics we have
\be
\langle p,q;\epsilon |\Psi\rangle = 
\int dx~\langle p,q;\epsilon |x\rangle
\psi (x)
\ee 
where $\psi (x) \equiv\langle x|\Psi\rangle$  is a wave function 
of coordinate representation,
$|p,q;\epsilon\rangle$ is a coherent state [10]
\be
{\bf \hat a} |p,q;\epsilon\rangle = 
\frac {q + i\epsilon p}{\sqrt {2\epsilon\hbar}}~ |p,q;\epsilon\rangle
\ee
and respectively, the ``matrix element" $\langle p,q;\epsilon |x \rangle $ 
is given by
$$
\langle p,q;\epsilon |x \rangle = 
\left (\frac {1}{\pi\epsilon\hbar}
\right )^{1/4}
\exp {(\frac {i}{2\hbar}pq)} \exp {(-\frac {i}{\hbar}px)}
\exp {(-\frac {(x-q)^2}{2\epsilon\hbar})}
\eqno (3.13')
$$
Then, from (3.12) and (3.13$'$) we obtain
\be
\lim \limits_{\epsilon \to 0}~\langle p,q;\epsilon |\Psi\rangle  
\left (\frac {1}{4\pi\epsilon\hbar}
\right )^{1/4} =
\exp{(-\frac{i pq}{2\hbar})}~\psi (q)
\ee
It is well known that the matrix element
$\langle p,q;\epsilon |\Psi\rangle$ 
defines the wave function of holomorphic representation (see [8, 10])
\be
\langle p,q;\epsilon |\Psi\rangle =
\exp{(-\frac {1}{2}|z|^2)}F (z^*)\equiv
\tilde \Psi_\epsilon (p, q) 
\ee
where the variables $p,q$ and $z,z^*$ are related by (3.10). On the other
hand, from the equivalence of holomorphic representation and E-quantization
scheme, the wave function  $\tilde \Psi_\epsilon (p, q)$ in (3.15) can be
considered as the vector of physical Hilbert space ${\cal H}_\epsilon$
(compare (3.11) and (3.15)).
Then, (3.12) and (3.14) will be similar to (3.4) and (3.6) 
respectively. Only, it should be noted, that the two physical states 
$\Psi_\epsilon (p,q)$ and $\tilde \Psi_\epsilon (p,q)$, 
constructed by the same function 
$\psi (q)\in {\cal L}_2({\cal R}^1)$, 
are different ($\Psi_\epsilon (p,q) \neq \tilde \Psi_\epsilon (p,q)$)
(see (3.4) and (3.12)), 
and they coincide only in the limit $\epsilon \rightarrow 0$. 
This short remark indicates different possibilities
of described limiting procedure (for more details see Appendix A).

\vspace{0.5cm}
\noindent
Example 2. Phase space is a cylinder 
${\cal M} \equiv
{\cal R}^1\times {\cal S}^1$
with the coordinates $\xi^1 \equiv S \in {\cal R}^1,~ 
\xi^2 \equiv \varphi \in S^1$ and the symplectic form
$\omega = dS\wedge d\varphi $. 
This is a model of rotator where $S$ is an angular momentum.

Since a cylinder is a cotangent bundle over a circle,
the canonical quantization for this model is realized on the space
of square integrable functions $\psi (\varphi )$ on a circle
($\psi (\varphi )\in {\cal L}_2({\cal S}^1)$). The quantization
rule (1.4) gives
\be
\hat S~ \psi (\varphi )=-i\hbar\partial_\varphi \psi (\varphi )~~~~
\hat {\cos\varphi}~\psi (\varphi )=\cos\varphi~\psi (\varphi )
~~~~~~~\hat {\sin\varphi}~\psi (\varphi )=\sin\varphi~ \psi (\varphi )
\ee
and the operator $\hat S$ has the discrete spectrum $S_n=n\hbar$, 
($n\in Z$), with the eigenfunctions
$\psi_n(\varphi )=1/{\sqrt {2\pi}}~\exp {(in\varphi )}$. 

The coordinate $\varphi $ is not  global, 
and for the 1-form we choose $\theta = Sd\varphi$.
The set of functions
\be
f_1=S,~~~~ f_2=\cos\varphi ,~~~~ f_3 =\sin\varphi 
\ee
is complete (with the relation $f_2^2 + f_3^2 = 1$), 
and for the corresponding constraint operators we get
$$
{\hat \Phi}_S =S + i\hbar\partial_\varphi ,~~~~~
\hat \Phi_{\cos \varphi} = i\hbar\sin\varphi \partial_S,~~~~~~
\hat \Phi_{\sin\varphi} = -i\hbar\cos\varphi \partial_S
\eqno (3.17')
$$
Note, that there is a possibility to have the
complete set with only two functions as well. For example,
\be
{\tilde f}_1 = e^{S/\lambda }\cos \varphi ~~~~~~
{\tilde f}_2 
= e^{S/\lambda }\sin \varphi
\ee
where $\lambda$ is some constant parameter (with dimension of angular momentum).
These functions
are global coordinates on a cylinder and 
they give the map of a cylinder
on to a plane without origin: $({\tilde f}_1, {\tilde f}_2) \in 
{\cal R}^2 - \{0\}$.

From (3.17$'$) we see that in the E-quantization scheme
the wave functions $\psi (\varphi )$ of ``$\varphi$ representation"
can be obtained by
\be
\hat \Phi_{\cos \varphi}~ \Psi_{ph}(S,\varphi )=0 ~~~~~\mbox {and}~~~~~~
\hat \Phi_{\sin \varphi}~ \Psi_{ph}(S,\varphi )=0 
\ee
But it is clear that these functions are not normalizable on the cylinder.
Situation with the condition
\be
\hat \Phi_S~ \Psi_{ph}(S,\varphi )=0 
\ee
is more complicated, since  equation (3.20) has no global regular
solutions. In the class of generalized functions one can find
the solutions of the type
\be
 \Psi_{ph,n}= \delta (S-n\hbar)\exp{(in\varphi)}~~~~~(n\in Z) 
\ee
which obviously are not square integrable on the cylinder.
To investigate these classes we need a limiting procedure 
as it was done for  Example 1. Such a procedure we consider in the next 
section with some motivation and generalization of condition (2.9), and 
here, in the rest part of this section, we construct some physical 
Hilbert spaces as the subspaces of
${\cal L}_2({\cal R}^1\times {\cal S}^1)$.
For this we introduce the complex coordinates related to (3.18)
\be
z= \tilde f_1 -i\tilde f_2 = \exp {(S/\lambda -i\varphi )}~~~~~~~~
z^*= \tilde f_1 +i\tilde f_2 = \exp {(S/\lambda +i\varphi )}
\ee
and impose condition like (2.9) for $\epsilon =1$
\be
\hat \Phi_{z^*}~|\Psi_{ph}\rangle =0
\ee
This is equivalent to the equation
\be
\left (\partial_z + \frac {\lambda}{2\hbar}\frac {\log |z|}{z}\right ) 
\Psi_{ph} (z,z^*) =0
\ee
and for the solutions we get
\be
\Psi_{ph} (z, z^*) = \exp {\left (-\frac {\lambda}{2\hbar}
(\log |z|)^2 \right )}
\psi(z^*)
\ee
where $\psi (z^*)$ is any holomorphic function ($\partial_z\psi =0$)
on the plane without origin and it has the expansion
$$
\psi (z^*) = \sum_{n=-\infty}^{\infty}~ d_n{z^*}^n
$$
Respectively, in ($S,\varphi $) coordinates (3.25) takes the form
\be
\Psi_{ph} (S,\varphi ) = \sum_{n=-\infty}^{\infty}~ c_n
\exp {\left (-\frac {(S-n\hbar)^2}{2\lambda\hbar}\right )}\exp {(in\varphi )}
\ee
with $c_n = d_n \exp {({\hbar n^2}/{2\lambda})}$, and square
integrability gives
\be
 \sum_{n=-\infty}^{\infty}~ |c_n|^2 < \infty
\ee
From (1.9) and (3.17$'$), the pre-quantization operator of angular momentum 
is $\hat {R}_S= -i\hbar\partial_\varphi$. It is a well defined operator
on the physical
subspace (3.26), and has the same non-degenerated spectrum, as
the operator $\hat S$ of the canonical quantization.   
Thus, we see the unitary equivalence of these two quantizations.

\vspace*{1cm}

\section{Minimal Fluctuations of Quantum Constraints}
\setcounter{equation}{0}
\noindent
For Example 1 of the previous section the constraint operators 
$\hat {\Phi}_p$ and $\hat {\Phi}_q$  have the canonical 
commutation relations (see (3.2))
\be
[\hat {\Phi}_p, \hat {\Phi}_q] =i\hbar
\ee
The condition  (3.3) is equivalent to the
choice of physical states $|\Psi_\epsilon\rangle$ as the ``vacuum" states
in $\Phi_p, \Phi_q$ variables
\footnote {Recall
that due to quantum uncertainties, we can not put 
$\hat {\Phi}_p |\Psi\rangle=0$ and $\hat {\Phi}_q |\Psi\rangle=0$
simultaneously.}. Then, the mean values of constraints
are equal to zero
$$
\langle \Psi_\epsilon|\hat {\Phi}_p|\Psi_\epsilon\rangle =0~~~~~~~~~~~~~~~
\langle \Psi_\epsilon|\hat {\Phi}_q|\Psi_\epsilon\rangle =0
\eqno {(4.1')}
$$
and the product of quadratic fluctuations is minimal
$$
\langle \Psi_\epsilon|\hat {\Phi}^2_p|\Psi_\epsilon\rangle 
\langle \Psi_\epsilon|\hat {\Phi}^2_q|\Psi_\epsilon\rangle =
\frac {\hbar^2}{4}
\eqno {(4.1'')}
$$
Thus, the meaning of the condition (2.9) for this simple
example is that the obtained physical states $|\Psi_\epsilon\rangle$
provide the best realization of the classical constraints
${\Phi}_p =0,~{\Phi}_q =0$ on the quantum level.

Let us consider the condition (2.9) in general case.
Note, that if two functions $f_a$ and $f_{N+a}$ are canonically
conjugated: $\{f_a,f_{N+a}\}=1$, then the
corresponding constraint operators have canonical commutation relations
(see (2.7$'$)). Therefore, for the construction of physical states 
by (2.9) it is natural to choose the function
$f_{N+a}$ as a canonically conjugated to $f_a$, and repeat calculations
of Example 1 in $f_a$, $f_{N+a}$ variables.
Unfortunately, this simple procedure, in general, fails. The reason is that
the canonically conjugated variable $f_{N+a}$ usually
exists only locally and
corresponding constraint $ {\Phi}_{f_{N+a}}$ is not well defined both
on classical and quantum levels.
For example, canonically conjugated variable to the harmonic 
oscillator Hamiltonian $H=1/2(p^2+q^2)$ is the polar angle $\alpha$
\be
p=\sqrt {2H}\cos\alpha ~~~~~~~~~~~~q=\sqrt {2H}\sin\alpha
\ee
Choosing the 1-form $\theta =1/2(pdq-qdp)$, we get 
$$
{\hat \Phi}_H =H + i\hbar\partial_\alpha
\eqno {(4.2')}
$$
and for the operator
$\hat {\Phi}_\alpha$ one can formally write
$\hat {\Phi}_\alpha =-i\hbar\partial_H$, but this operator is not
self-adjoint . Then, though the equation 
$$
(\hat {\Phi}_H+i\epsilon\hat {\Phi}_\alpha)|\Psi\rangle=0
\eqno {(4.2'')}
$$
has integrable solutions (for example $\Psi (p,q) = 
\exp {(- {H^2/{2\epsilon\hbar}})}$), nevertheless they are not acceptable for
the physical states, since the mean values of the constraint
operators $\hat {\Phi}_H$ and $\Phi_\alpha$ do not vanish, and  
minimization of quadratic fluctuations is not achieved as well.

For $\epsilon =0$ one can write the formal solution of (4.2$''$) (like (3.21)):
$\Psi =\delta (H-\hbar n)\exp {(in\alpha)}$, and since $H\geq 0$,
such ``solutions" are non-zero only for $n\geq 0$. Then, 
the pre-quantization operator $\hat {R}_H=-i\hbar\partial_\alpha$
has the spectrum $H_n=\hbar n, n\geq 0$.
The situation is similar for any completely integrable system [11].
In action-angle variables $I_a, \varphi_a~ (a=1,...,N)$
we have  the 1-form $\theta =I_ad\varphi_a$ and the Hamiltonian
$H=H(I_1,...,I_N)$.
Then, the constraint and pre-quantization operators take the form
\be
{\hat \Phi}_{I_a} =
I_a + i\hbar\partial_{\varphi_a}
\ee
$$ 
{\hat R}_{I_a} =-i\hbar\partial_{\varphi_a}~~~~~~~~~
{\hat R}_{H} =H-\frac{\partial H}{\partial I_a}{\hat \Phi}_{I_a}
\eqno {(4.3')}
$$
If $\varphi_a$ are the cyclic variables ($\varphi_a\in {\cal S}^1$)
\footnote {Note, that operator $\hat {\Phi}_{\varphi_a}$
is Hermitian, when the corresponding action variable is unbounded
$-\infty \leq I_a \leq \infty$ (as the angular momentum $S$ for
the  Example 2).},
then by described formal operations we obtain 
the ``physical states"
\be
\Psi_{ph}(I,\varphi ) =
\prod_{a=1}^{N}\delta (I_a-\hbar n_a)\exp {(in_a\varphi_a)}
\ee
as the ``solutions" of the equations
$$
\hat {\Phi}_{I_a}\Psi_{ph}(I,\varphi )=0
\eqno {(4.4')}
$$
The spectra of pre-quantization operators (4.3$'$) on these
``physical states" are 
$$(I_a)_{n_a}=\hbar n_a~~~~~~~
{\mbox {and}}~~~~~~ H_{n_1,...,n_N}=H(\hbar n_1,...,\hbar n_N)
$$ 
where $n_a$ are integer numbers, 
and corresponding admissible values are chosen according to the possible 
classical values of the variables $I_a$
(as, for example, $n\geq0$ for the harmonic oscillator).
It is remarkable, that these formal 
results  correspond to the quantization rule
\be
I_a\Delta\varphi_a =\oint p_adq_a =2\pi\hbar n_a
\ee
which is almost the semi-classical one. From these formal
operations it seems that the quantum
problem is solvable for any completely integrable system; but of course, 
all these expressions here have only symbolic meaning
and (4.4) needs further specification, taking account of 
$N$ other constraints and limiting procedure as well.

After these remarks let us consider
the case when observables $f_a$ and $f_{N+a}$ (in (2.9)) 
are not canonically
conjugated to each other. For the convenience we use the notations
$f_a\equiv f$, $f_{N+a}\equiv g$ and introduce corresponding constraint 
operators
$\hat {\Phi}_f$ 
and
$\hat {\Phi}_g$.

It turns out, that in general, equation (2.9) has no normalizable
solutions at all, and choice of sign (or value) of $\epsilon$ does not
help\footnote {Sometimes, even normalizable solutions are not acceptable
as well (see (4.2) and Section 5)}. 
For example, if $f$ is a kinetic energy $f={p^2}/2m$,
and $g$ is a coordinate $g=q$ of one dimensional system, 
then  (2.9) takes the form (with $\theta=pdq$ and $m=1$)
$$
(p^2+i\hbar p\partial_q+\epsilon\hbar\partial_p)
\Psi_\epsilon (p,q)=0
$$
and the solutions
$$
\Psi_\epsilon (p,q)= \exp {(-\frac {\epsilon p^3}{3\epsilon\hbar})}
\psi (p^2+2i\epsilon q)
$$
evidently are not normalizable. Of course, for this example we can return to 
the canonical coordinates $p,q$ and make reduction (3.3) with
constraints $\Phi_p$ and $\Phi_q$; but if we intend to deal with
arbitrary observables and symplectic manifolds,
we have to generalize the condition (2.9). 
For this we introduce
the minimization principle for quadratic fluctuations.

Quadratic fluctuations of two Hermitian operators 
$\hat {\Phi}_f$ 
and
$\hat {\Phi}_g$ can be characterized by the functional $U(\Psi )$
\be
U(\Psi )\equiv
\langle \Psi|\hat {\Phi}_f^2|\Psi\rangle 
\langle \Psi|\hat {\Phi}_g^2|\Psi\rangle 
\ee
where $|\Psi\rangle$ is a vector with the unit norm 
$\langle \Psi|\Psi\rangle =1$.

Then, one can postulate the principle that the physical states 
provide minimization of this functional (see (4.11$'$)). 
For two arbitrary Hermitian operators minimization problem
of uncertainties was studied in [12], 
and in Appendix C we present some results
of this investigation.
Note, that in [12] the minimization problem was 
considered for another functional $U_1(\Psi )$
\be
U_1(\Psi )\equiv\frac{\langle\Psi|\hat {\Phi}^2_f|\Psi\rangle
\langle\Psi|\hat {\Phi}^2_g|\Psi\rangle}
{\langle\Psi|\hat A|\Psi\rangle^2}
\ee
as well. Here the operator $\hat A$ is the commutator 
\be
\hat A =-\frac{i}{\hbar}[\hat {\Phi}_f, \hat {\Phi}_g] 
\ee
and only for the $c$-number operator the functionals 
$U(\Psi )$ and $U_1(\Psi )$ are equivalent.
In this section we consider only the 
functional $U(\Psi )$.

Then, using results of [12] (see (C.4) and (C.5)), 
minimization principle gives that  
the physical wave functions
$|\Psi_{ph}\rangle$ can be obtained from the equation 
\be
\frac{1}{2a^2}\hat {\Phi}_f^2|\Psi_{ph}\rangle+\frac{1}{2b^2}
\hat {\Phi}_g^2|\Psi_{ph}\rangle=|\Psi_{ph}\rangle
\ee
and subsidiary conditions 
\be
a^2= \langle \Psi_{ph}|\hat {\Phi}_f^2|\Psi_{ph}\rangle
~~~~~~~~~b^2= \langle \Psi_{ph}|\hat {\Phi}_g^2|\Psi_{ph}\rangle
\ee
where $a$ and $b$ are some fixed parameters. Possible values 
of these parameters are defined from the following procedure: 
At first we have to
solve the equation (4.9) with free parameters $a, b$ and select the
solutions with unit norm which satisfy (4.10). Usually after this
we still have a freedom in $a$ and $b$. Then we must choose one of 
those pairs with minimal product of $ab$ (we assume
both $a$ and $b$ to be nonnegative). The fixed values of the parameters
$a$ and $b$ provide that the solutions of (4.9) form the linear space
as the subspace of ${\cal L}_2({\cal M})$.
This subspace should define the physical Hilbert space 
${\cal H}_{ph}\equiv{\cal H}_{(a,b)}$ of the system.

Thus, instead of the first order differential equation (2.9) with one parameter
$\epsilon $ (see (2.9)) we get the second order equation (4.9)
with two parameters $a$, $b$ and subsidiary conditions (4.10).
Note , that possible limiting procedure in (4.9) for 
$a\rightarrow 0$ (or $b\rightarrow 0$ )  can specify 
the physical states
$|\Psi_{ph}\rangle$ with $\hat {\Phi}_f|\Psi_{ph}\rangle =0$ 
(or $\hat {\Phi}_g|\Psi_{ph}\rangle =0$). 

For the test of formulated principle, at first we consider 
again  Example 1.
In this case the constraint operators
$\hat {\Phi}_f \equiv\hat {\Phi}_p$ and $\hat {\Phi}_g \equiv\hat {\Phi}_q$  
have the canonical commutation relations (4.1$'$).
Then, (4.9) looks like the harmonic oscillator eigenvalue problem
with the frequency $\omega =1/{ab}$ and the eigenvalue $E=1$.
Respectively, we get $\hbar(n+1/2)=ab$. 
One can check, that all the oscillator's eigenstates $|n\rangle$
satisfy the conditions (4.10), and
therefore the minimal $ab$ ($ab =\hbar/2$)
corresponds to the vacuum state ($n=0$) given by
$(a\hat {\Phi}_q -ib\hat {\Phi}_p)|\Psi_{ph}\rangle =0 $.
Thus, for the physical states we arrive again to (3.3) with $\epsilon =
b/a$, and the limiting procedure 
$a\rightarrow 0$ (or $b\rightarrow 0$ ) 
can can be accomplished in a similar way.

Now, let us consider  Example 2 with the constraint operators (3.17$'$).
For the convenience we can construct the operator
$\hat O \equiv\hat {\Phi}^2_{\sin\varphi}+
\hat {\Phi}^2_{\cos\varphi}$, 
and  minimize
the product $\langle\Psi |\hat {\Phi}_S^2|\Psi\rangle 
\langle\Psi |\hat O|\Psi\rangle$.
From (3.17$'$) we have $\hat O = -\hbar^2\partial^2_S$, and we see that this
operator is a square of the Hermitian operator 
$\hat {\Phi}_\varphi \equiv 
-i\hbar\partial_S$ ($\hat {O} =\hat {\Phi}_\varphi^2$). 
Then, from the variation principle we get
the equation (4.9) with $\hat {\Phi}_f = S+i\hbar\partial_\varphi$
and $\hat {\Phi}_g = -i\hbar\partial_S$. Since these two Hermitian operators
have canonical commutation relations, we 
arrive again at the oscillator problem.
Only, now the ``ground" state should be obtained from the equation
\be
(S+i\hbar\partial_\varphi +\frac{a}{b}\hbar\partial_S)|\Psi_{ph}\rangle 
\ee
Hence, for this example, using the minimization principle,
we arrive at the equation (4.11). It is interesting to note,
that the equations (4.11) and (3.24) are equivalent, and
the functions (3.26) with $\lambda = a/b$ are the 
solutions of (4.11).
Indeed, one can check that (4.11) can be obtained
from (3.24) by multiplication on $2\hbar z$ (see (3.22), (3.24)).

In (4.11) we can accomplish the limiting procedure to the 
equations (3.20) (or (3.19)) taking corresponding limits
$a/b\equiv\lambda \rightarrow 0$
(or $\lambda \rightarrow \infty$). 

From  (3.26) we see that the functions
\be
\Psi_{\lambda ,n} (S,\varphi ) = 
\left (\frac {\hbar}{\pi\lambda}\right )^{1/4}
\exp {\left (-\frac {(S-n\hbar)^2}{2\lambda\hbar}\right )}\exp {(in\varphi )}
\ee
form the basis for the physical states (4.11). This basis satisfies the 
following ortho-normality conditions
$$
\langle \Psi_{\lambda ,n}|\Psi_{\lambda ,m}\rangle \equiv
\int \frac {dSd\varphi}{2\pi\hbar}
\Psi^*_{\lambda ,n} (S,\varphi ) \Psi_{\lambda ,m} (S,\varphi ) = \delta_{nm}
$$
With suitable normalization these basis functions
have the limits as
 $\lambda \rightarrow 0$ 
(or $\lambda \rightarrow \infty$)
in the dual space ${\cal L}^*_2({\cal R}^1\times {\cal S}^1)$ (see Appendix A). 
Indeed, the limit $\lambda \rightarrow 0$ of the function
$$
\tilde {\Psi}_{\lambda ,n} (S,\varphi ) = \frac {1}{\sqrt{2\hbar}}
\left (\frac {1}{\pi\lambda}\right )^{1/4}
\Psi_{\lambda ,n} (S,\varphi ) 
$$
is the generalized function (3.21) which is a well
defined linear functional on
${\cal L}_2({\cal R}^1\times {\cal S}^1)$. According to the rule (2.11)
physical states (3.21) with different $n$ form the ortho-normal basis of 
the corresponding reduced Hilbert space.
Similarly, we can take the limit 
$\lambda \rightarrow \infty$ for the functions
 $$
\tilde {\tilde {\Psi}}_{\lambda ,n} (S,\varphi ) = \frac {1}{\sqrt{2}}
\left (\frac {\lambda}{\pi\hbar}\right )^{1/4}
\Psi_{\lambda ,n} (S,\varphi ) 
$$
and obtain
$$
\lim \limits_{\lambda \to 0}
\tilde {\tilde {\Psi}}_{\lambda ,n} (S,\varphi ) = 
\psi_n(\varphi )=1/{\sqrt {2\pi}}~\exp {(in\varphi )}
$$
This is the basis of the Hilbert space of canonical quantization (see (3.16)),
and we have the same ortho-normality conditions due to the rule (2.11).

Obtained physical wave functions have other remarkable properties
with respect to the described limiting procedure.

Let $\Psi_{\lambda} (S,\varphi )$ be any physical state (4.11)
with unit norm. Then,
\be
\Psi_{\lambda} (S,\varphi ) = 
\sum_{n=-\infty}^{\infty}~ c_n
\Psi_{\lambda ,n} (S,\varphi ) 
~~~~~~\mbox {with}~~~~~~~ 
\sum_{n=-\infty}^{\infty}~ |c_n|^2 =1
\ee
where $\Psi_{\lambda ,n} (S,\varphi )$ is the basis (4.12).
If we integrate by $\varphi$ the square of modulus
of this function, and then take the limit $\lambda \rightarrow 0$,
we obtain
\be
\lim \limits_{\lambda \to 0}
\int \frac {d\varphi}{2\pi\hbar}~|\Psi_{\lambda} (S,\varphi )|^2 
=\sum_{n=-\infty}^{\infty}~ |c_n|^2 \delta (S-n\hbar)
\ee
We see that the right hand side of (4.14) describes the distribution
function for the measurement of angular momentum $S$ in
the state $\Psi_\lambda $.

The same we can obtain for the normalized physical states (3.12)
of Example 1. Namely,
\be
\lim \limits_{\epsilon \to 0}
\int \frac {dp}{2\pi\hbar}~|\langle p,q;\epsilon|\Psi\rangle |^2 
=|\psi (q)|^2
\ee
where we use the representations (3.12) and (3.13$'$).

It is interesting to note that the integrands
in (4.14) and (4.15) have similar properties. Indeed, using that
$$
\lim \limits_{\lambda \to 0}
\Psi^*_{\lambda ,n} (S,\varphi )  
\Psi_{\lambda ,m} (S,\varphi ) = 0,~~~~~{\mbox {when}}~~~~~~m\neq n
\eqno (4.14')
$$
we get
$$
\lim \limits_{\lambda \to 0}
|\Psi_{\lambda} (S,\varphi )|^2 =
2\pi\hbar\sum_{n=-\infty}^{\infty}~ |c_n|^2 \delta (S-n\hbar)
$$
For Example 1, of course the integrand in (4.15) has zero limit
(when $\epsilon \rightarrow 0$),
since it is integrable on the plane and in this limit it does not depend on 
momentum $p$. If we neglect this zero factor we get the coordinate
distribution function $|\psi (q)|^2$ (see 3.14).

These properties we use for the physical interpretation
of wave functions $\Psi_{ph}$ in Section 6, and now we
return to the conditions (2.9) and minimization of $U_1(\Psi)$
for further investigation.

\vspace*{1cm}

\section{Minimal Uncertainties and Coherent States }
\setcounter{equation}{0}
\noindent
We can consider the minimization principle for quadratic fluctuations
using the functional $U_1(\Psi )$ (see (4.7)) as well. In this case instead
of (4.9) we get the equation (see (C.10))
\be
\frac{1}{2a^2}\hat {\Phi}_f^2|\Psi_{ph}\rangle +\frac{1}{2b^2}
\hat {\Phi}_g^2|\Psi_{ph}\rangle -\frac 
{\hat A}{A}|\Psi_{ph}\rangle =0
\ee
where ${\hat A}$ is a commutator (4.7), $A$ is a parameter, and solutions
$|\Psi_{ph}\rangle$ should satisfy (4.10) and the condition
$\langle\Psi_{ph}|{\hat A}|\Psi_{ph}\rangle=A$ as well (see Appendix C).

There is some relation between the minimization of the functional
$U_1(\Psi )$ and the condition (2.9).
In our notations the condition (2.9) has the form
\be
(\hat {\Phi}_f +i\epsilon\hat {\Phi}_g )|\Psi_\epsilon\rangle=0
\ee
and for the wave function $\Psi_\epsilon(\xi )$ this is the 
first order differential equation.
Of course, it is much easier to analyze solutions of (5.2)
\footnote {Practically it is always integrable.},
then to investigate (4.9) (or (5.1)),
which are the second order equations with two (or three) free parameters
and subsidiary conditions (4.10). But, to be acceptable for the physical
states, the corresponding solutions of (5.2) should belong to the domain of 
definition of self-adjoint operators
$\hat {\Phi}_f$ and $\hat {\Phi}_g$.
Except finiteness of the norm of $|\Psi_\epsilon\rangle$, this means that the 
operators $\hat {\Phi}_f$ and $\hat {\Phi}_g$
must be Hermitian on these functions.
As it was pointed out,
in general, these conditions are not fulfilled, and 
in that case we have to use
the minimization
principle for quadratic fluctuations of quantum constraints.
But, if for some real $\epsilon $, 
solutions of (5.2) satisfy the two conditions mentioned above,
then we have (see Appendix C)
$$
\langle \Psi_\epsilon |\hat {\Phi}_f^2|\Psi_\epsilon \rangle =
\frac{\hbar\epsilon A}{2}~~~~~
\langle \Psi_\epsilon |\hat {\Phi}_g^2|\Psi_\epsilon \rangle =
\frac{\hbar A}{2\epsilon}~~~~~
{\mbox {where}} 
~~~~\langle\Psi_\epsilon |{\hat A}|
\Psi_\epsilon \rangle =A
$$
and corresponding physical states $|\Psi_\epsilon\rangle$ 
provide minimization of the functional
$U_1(\Psi )$: $U_1(\Psi_\epsilon )=\hbar^2/4$. 
Note (and it is natural) that
such  functions $|\Psi_\epsilon\rangle$ satisfy (5.1) 
($|\Psi_{ph}\rangle=|\Psi_\epsilon\rangle$), with 
$a^2={\hbar \epsilon A}/{2}$,
$b^2={\hbar A}/{2\epsilon}$ and
$A=\langle\Psi_\epsilon |{\hat A}|
\Psi_\epsilon \rangle$. To be convinced, it is sufficient to act on (5.2)
by the operator
$\hat {\Phi}_f -i\epsilon\hat 
{\Phi}_g $.

When the commutator $\hat A$ in (5.1) is a $c$-number, then (5.1)
and (4.9) are equivalent and they define the same physical
Hilbert spaces as the subspaces of ${\cal L}_2({\cal M})$.
But, in general, for given observables $f$ and $g$
these subspaces are different
and to understand which one is more suitable
further investigation is required.
On the other hand, the functionals $U$ and $U_1$ (and corresponding
reduced physical Hilbert spaces) essentially depend on the choice
of the pair of observables $f, g$. It turns out that reduced physical
Hilbert spaces obtained by minimization of the functionals $U$ and $U_1$
can be the same, even if the pair of observables $f, g$ for $U$ and $U_1$ 
are different.
For example, physical states (3.25)-(3.26) were obtained 
from (5.2) with
$f\equiv \tilde f_1,~g\equiv \tilde f_a$ and $\epsilon =1$ (see (3.22)-(3.23)). 
Respectively, these solutions minimize the functional 
$U_1(\Psi )$.
In section 4 it was checked, that the same physical states minimize
$U(\Psi )$ 
as well, but for $U(\Psi )$ the functions $f$ and $g$ are different
($f\equiv S$, $g\equiv \varphi $ (see (4.11)).

Let us return again to the choice of physical states by condition (2.9)
(or (5.2)).
For simplicity we consider the two dimensional case.

Suppose that solutions of (5.2) for some real $\epsilon$
satisfy the two required conditions, and hence, they are
acceptable for the physical states.
In complex variables $z=f-i\epsilon g,~z^* =f+i\epsilon g$ 
condition (5.2) can be written as $
\hat {\Phi}_{z^*}
|\Psi_{ph}\rangle=0$.
The corresponding differential equation  has the form (see (3.10$'$)
and (3.24))
\be
\left (\partial_z-\frac{i}{\hbar}\theta_z(z,z^*)\right )\Psi_{ph}(z,z^*)=0
\ee
where $\theta_z$ is the component of the 1-form $\theta =\theta_zdz
+\theta_{z^*}dz^*$. Solutions of (5.3) are
\be
\Psi_{ph}(z,z^*)
= \exp {(-\frac{1}{2}S(z,z^*))}\psi (z^*)
\ee
with arbitrary\footnote {Class of holomorphic functions $\psi (z^*)$ 
should provide a finite norm of physical
states $\Psi_{ph}(z,z^*)$}
$\psi (z^*)$, and $S={{2i}/\hbar}\int dz \theta_z$. These functions
define the physical Hilbert space ${\cal H}_\epsilon$.

The pre-quantization operator $\hat R_{z^*} =z^*-\hat {\Phi}_{z^*}$
acts invariantly on the physical states (5.4), and this action is given
by the multiplication of corresponding wave functions  
$\psi (z^*)$ by $z^*$
\be
\hat R_{z^*} \Psi_{ph}(z,z^*) =z^*\Psi_{ph}(z,z^*)
\ee
From (2.7$'$) and (2.5) we have 
$$[\hat R_z,\hat \Phi_{z^*}]=-i\hbar 
\hat \Phi_{\{z,z^*\}}=2\epsilon\hbar \partial_z(\{f,g\}) 
\hat {\Phi}_z +2\epsilon\hbar \partial_{z^*}(\{f,g\}) \hat {\Phi}_{z^*}
$$ 
and if the Poisson bracket $\{f,g\}$ is not a constant, 
then the Physical Hilbert space (5.4) is not invariant
under the action of pre-quantization operator $\hat R_z$.
In this case, the deformation
procedure is problematic (see Appendix B), and to define the operator 
$\hat z$ we use the relation between $z$, $z^*$ variables.
Since the operator $\hat z^+ \equiv \hat R_{z^*}$ is well
defined on the physical states (5.4), it is natural to 
introduce the operator $\hat z$ as Hermitian conjugated to  
$\hat R_{z^*}$: $\hat z\equiv(\hat R_{z^*})^+$.
Respectively, operators $\hat f$ and $\hat g$ will be
\be
\hat f =\frac{1}{2}(\hat z +\hat z^+)
~~~~~~~~\hat g =\frac{i}{2\epsilon}(\hat z -\hat z^+)
\ee
If $
\Psi_{ph,n}(z,z^*)$ is some ortho-normal basis of the physical
Hilbert space (5.4), then the action of the operator $
\hat z$
on any state 
$\Psi_{ph}(z,z^*)$ can be written as
\bea
\hat z\Psi_{ph}(z,z^*)&=&
\sum_{n}\Psi_{ph,n}(z,z^*)\langle\Psi_{ph,n}|\hat z|\Psi_{ph}\rangle
=\sum_{n}\Psi_{ph,n}(z,z^*)\langle\hat z^+\Psi_{ph,n}|\Psi_{ph}\rangle
=\nn \\
&=&\sum_{n}\Psi_{ph,n}(z,z^*)\int d\mu~ 
\Psi^*_{ph,n}(z',z'^*)z'\Psi_{ph}(z',z'^*)
\eea
where $ d\mu \equiv d\mu (z',z'^*)$ is the standard measure (1.12).

Let us introduce the wave function $
\chi_\zeta (z,z^*)$:
\be
\chi_\zeta (z,z^*) \equiv 
\sum_{n}\Psi^*_{ph,n}(\zeta ,\zeta^* )\Psi_{ph,n}(z,z^*)
\ee
Here $\zeta$ is considered as a complex parameter, and can take 
values in the same domain as the variable $z$. So, (5.8) is an
expansion of the wave function $\chi_\zeta (z,z^*)$ in the basis
$\Psi_{ph,n}(z,z^*)$ with coefficients
$\Psi^*_{ph,n}(\zeta ,\zeta^*)$.

With some assumptions about the analytical structure on $\cal M$ one can prove
(see [10] and [13]), that the function
$\chi_\zeta (z,z^*)$ is well defined, 
it is square integrable
$$
\int d\mu ~
|\chi_\zeta (z',z'^*)|^2 <\infty
$$
and the corresponding norm
$$
\rho^2_\zeta \equiv \int d\mu~ |\chi_\zeta |^2 =
\sum_{n}\Psi^*_{ph,n}(\zeta ,\zeta^* )\Psi_{ph,n}(\zeta ,\zeta^*)
= \chi_\zeta (\zeta ,\zeta^*)
\eqno (5.8')
$$
does not depend on the choice of the basis 
$\Psi_{ph,n}(\zeta ,\zeta^*)$.

Then, for an arbitrary physical state 
$|\Psi_{ph}\rangle$, (5.8) yields
\be
\langle
\chi_z| 
\Psi_{ph}\rangle=\int d\mu~ 
\chi^*_z (z',z'^*) 
\Psi_{ph}(z',z'^*)
=\Psi_{ph}(z,z^*)
\ee

If we act with the operator $\hat z$ on the state
$\chi_\zeta(z,z^*)$, and use (5.9) and (5.5), we obtain
\bea
\hat z \chi_\zeta (z,z^*) =
\langle\chi_z|\hat z| \chi_\zeta\rangle=
\langle\chi_\zeta|\hat z^+| \chi_z\rangle^*=
(\hat z^+ \chi_z(\zeta ,\zeta^*))^*= \nn \\ 
(\zeta^* \chi_z(\zeta ,\zeta^*))^*=
\zeta~ \chi_\zeta(z ,z^*)
\eea
where in the last stage we take into account that
$$
 \chi^*_z(\zeta ,\zeta^*)= 
\chi_\zeta(z ,z^*)
$$
which is apparent from the definition (5.8). 

Thus, we see that
the function
$\chi_\zeta(z ,z^*)$ is the eigenstate of the operator $\hat z$
with the eigenvalue $\zeta $. 
This state is uniquely characterized by the complex parameter $\zeta$. 
Sometimes we omit the coordinates of the phase space as the arguments
of corresponding functions, and denote the state $\chi_\zeta (z, z^* )$
by $\chi_\zeta$, or $|\zeta\rangle$. We use also the notation
$|\zeta\rangle\equiv|\bar f,\bar g;\epsilon\rangle$, 
where $\bar f$ and $\bar g$ are the real and imaginary parts
of the complex number $\zeta$ respectively. From (5.8-10)
we have the following properties of $|\zeta \rangle$ states
\be
\int d\mu (\zeta )~  |\zeta \rangle \langle \zeta | = \hat I
\ee
$$
\langle z|\zeta \rangle =\chi_\zeta(z ,z^*)
\eqno (5.11')
$$
$$
\hat z|\zeta \rangle=\zeta|\zeta \rangle
\eqno (5.11'')
$$
It is remarkable, that the condition of completeness  (5.11) allows us
to introduce covariant and contravariant symbols of Berezin quantization
[13].

Further, for the Hermitian operators (5.6) the relation (5.11$''$)
takes the form
\be
(\hat f -i\epsilon\hat g)|\bar f,\bar g;\epsilon \rangle =
(\bar f-i\epsilon\bar g)|\bar f,\bar g;\epsilon
 \rangle
\ee
Then, we immediately get that 
\be
 \langle\bar f,\bar g;\epsilon|\hat f |\bar f,\bar g;\epsilon \rangle =\bar f
~~~~~~~ \langle\bar f,\bar g;\epsilon
|\hat g |\bar f,\bar g;\epsilon \rangle = \bar g
\ee
and using the method described in Appendix C (for the details
see [12]) we obtain
\be
\frac{ \langle\bar f,\bar g;\epsilon|(\hat f -\bar f)^2|\bar f,\bar g;\epsilon 
\rangle 
 \langle\bar f,\bar g;\epsilon|(\hat g -\bar g)^2|\bar f,\bar g;\epsilon 
\rangle }
{\langle\bar f,\bar g;\epsilon|\hat C|\bar f,\bar g;\epsilon \rangle^2 }=
\frac{\hbar^2}{4}
\ee 
where $\hat C$ is the commutator
 $\hat C =i/\hbar~[\hat f,\hat g]$. Note, that the operators
$\hat f$ and  $\hat g$ generally are not the pre-quantization ones, 
and respectively,
commutator $\hat C$ is not of the form (1.18). 

Thus, the quantum state 
$ |\bar f,\bar g;\epsilon \rangle$ minimizes
the quadratic fluctuations of the observables $f$ and $g$ around the values
$\bar f$ and $\bar g$. In this respect they are very similar 
to the coherent states
$|p,q,\epsilon\rangle$  (see (3.13)) which minimize the coordinate-momentum
uncertainty.

For the considered examples (see Section 3) many technical calculations with
coherent states can be accomplished explicitly.
In case of plane the ortho-normal basis for the physical states
(3.11) can be chosen as
\be
\Psi_{ph,n}(z,z^*)=\exp{(-\frac{1}{2}|z|^2)}\frac{{z^*}^n}{\sqrt {n!}}
\ee
Then, from (5.8) we get
\be
 \chi_\zeta(z ,z^*) =\exp{(-\frac{1}{2}|z|^2)}
\exp{(-\frac{1}{2}|\zeta|^2)}
\exp{(z^*\zeta)}
\ee
and since $ \chi_\zeta(\zeta ,\zeta^*)=1$, 
these states have the unit norm for arbitrary $\zeta$
(see (5.8$'$)). 
Comparing (5.11$''$) and (5.12) to (3.12) and (3.13)
we see, that the states $|\zeta\rangle$ are just 
the usual coherent states $|p,q,\epsilon\rangle$ mentioned above. 

For Example 2 we have
$$f= e^{S/\lambda }\cos \varphi ~~~~~~
g= e^{S/\lambda }\sin \varphi~~~~~\epsilon =1
$$
and the complex variables $z$ and $z^*$ are given by  (3.22).
The physical Hilbert space 
is defined by (3.25), or (3.26), and we have the ortho-normal basis (4.12). 
Here, we omit the index ``ph", arguments of the functions, and denote
the corresponding
basis by $\Psi_n$. The functions $\Psi_n$ are the eigenstates of the
operator $\hat S\equiv\hat R_S=-i\hbar\partial_\varphi$, with
eigenvalues $n\hbar$. Then, from (5.8) and (4.12), 
for the states $\chi_z$ ($z= 
\exp {(S/\lambda -i\varphi )}$) we get
\be
\chi_z = \left (\frac {\hbar}{\pi\lambda}\right )^{1/4}
\sum_{n=-\infty}^{\infty}~ 
\exp {\left (-\frac {(S-n\hbar)^2}{2\lambda\hbar}\right )}
\exp {(in\varphi )}~
\Psi_n
\ee 
and this state has the norm 
\be
||\chi_z ||^2=\left (\frac {\hbar}{\pi\lambda}\right )^{1/2}
\sum_{n=-\infty}^{\infty}~ 
\exp {\left (-\frac {(S-n\hbar)^2}{\lambda\hbar}\right )}
\ee
which is obviously finite. In the limit $\lambda \rightarrow 0$
we obtain
$$
||\chi_z ||^2 \rightarrow \sum_{n=-\infty}^{\infty}~ \delta (S/\hbar~-n)
\eqno (5.18')
$$
Let us introduce the operators $\hat V_\pm$ 
\be
\hat V_\pm \Psi_n=\Psi_{n\pm 1}
\ee
It is clear that the operators $V_\pm$ are equivalent to the phase operators
$\exp {(\pm in\varphi )}$ for
the canonical quantization. From the definition (5.19)
we have
\be
\hat V_+\hat V_-=\hat I=\hat V_-\hat V_+
~~~~~~\hat V_+^+=\hat V_-
\ee
Since the operator $\hat z^+$ acts as the multiplication on $z^*$,
for the basis vectors (4.12) we get
$$
\hat z^+\Psi_n=\exp{\left (\frac{\hbar n}{\lambda}+\frac{\hbar}{2\lambda}
\right )}\Psi_{n+1}
$$
Then, using the operator $\hat V_+$, we can represent the operator
$\hat z^+$ in two different forms
\be
\hat z^+=\exp{\left (\frac{\hat S}{\lambda}-\frac{\hbar}{2\lambda}\right )}
\hat V_+
=\hat V_+\exp{\left (\frac{\hat S}{\lambda}+\frac{\hbar}{2\lambda}\right )}
\ee
where we use that the basis vectors $\Psi_n$ are the eigenvectors
of the operator $\hat S$ with the eigenvalue $\hbar n$. 
Respectively, the Hermitian conjugated operator 
$\hat z$ is
\be
\hat z=\hat V_-\exp{\left (\frac{\hat S}{\lambda}-\frac{\hbar}{2\lambda}
\right )}
=\exp{\left (\frac{\hat S}{\lambda}+\frac{\hbar}{2\lambda}\right )}
\hat V_-
\ee
and, using (5.20), we obtain the commutator 
\be
[\hat z,\hat z^+]=2\exp{({2\hat S}/\lambda )}\sinh {(\hbar/\lambda )}
\ee
Note, that the corresponding classical commutation relation is
\be
\{\hat z,\hat z^+\}=\frac{2i}{\lambda}
\exp{(2 S/\lambda )}
\ee

Now, from (5.22) and (5.17), we can check that the states $\chi_z$
are the eigenstates of the operator $\hat z$ with the eigenvalues 
$z=\exp{(S/\lambda~-i\varphi)}$.

The states $\chi_z$ in (5.17) are defined for arbitrary value
of the variable $S$. At the same time, the states with fixed value
of the angular momentum ($\Delta S=0$) exist only for the discrete values
of $S$ ($S=\hbar n$). Of course, the states $\chi_z$ are not the eigenstates
of the operator $\hat S$, but, from (5.14), it is expected that 
$\Delta S \rightarrow 0$, when $\lambda \rightarrow 0$.
Therefore, it is interesting to investigate the behavior of the
states $\chi_z$, when $\lambda \rightarrow 0$.

Note, that expansion (5.17) can be considered as the definition
of the states $\chi_z$ for a quantum theory of a rotator in abstract
Hilbert space; only the basis vectors $\Psi_n$ should be the eigenstates
of the angular momentum operator $\hat S$ with the eigenvalues $S_n=\hbar n$.
With this remark we can neglect the dependence on the parameter
$\lambda$ in the basis vectors $\Psi_n$, and consider behavior
(when $\lambda \rightarrow 0$) of corresponding coefficients only.
If we introduce the vector $|S,\varphi;\lambda\rangle$ with unit norm
\be
|S,\varphi;\lambda\rangle \equiv \frac{\chi_z}{||\chi_z||}
\ee
then, from (5.17), we get
$$
|S,\varphi;\lambda\rangle =
\sum_{n=-\infty}^{\infty}~ 
\frac{d_n(S,\lambda)}{d(S,\lambda )}
\exp {(in\varphi )}~
\Psi_n
$$ 
where
$$
d_n(S,\lambda )=\exp {\left (-\frac {(S-n\hbar)^2}{2\lambda\hbar}\right )}
~~~~~~~d^2(S,\lambda )=\sum_{n=-\infty}^{\infty} d_n^2  ~~~~~(d>0)
$$
In the limit $\lambda \rightarrow 0$, 
$d_n(S,\lambda )/d(S,\lambda )~~\rightarrow c_n(S)$, and for the 
coefficients $c_n(S)$ we get:\\
a. $c_n(S)=0$, if $S<\hbar (n-1/2)$, or $S>\hbar(n+1/2)$;\\
b. $c_n(S)=1/{\sqrt {2}}$, if $S=\hbar (n-1/2)$, or $S=\hbar(n+1/2)$;\\
c. $c_n(S)=1$, if $\hbar (n-1/2)<S<\hbar(n+1/2)$.\\
From this we obtain, that $|S,\varphi;\lambda\rangle \rightarrow 
\exp {(in\varphi )}~\Psi_n$,
where $n$ is the nearest integer number to $S/\hbar$. But if
$S/{\hbar}$ is exactly in the middle of two integers: 
$S/{\hbar}=n+1/2$, then 
$|S,\varphi;\lambda\rangle \rightarrow 
1/\sqrt {2}~\left (
\exp {(in\varphi )}\Psi_n+\exp{(i(n+1)\varphi)}\Psi_{n+1}\right ) $.
So, when $\lambda \rightarrow 0$, all states 
$|S,\varphi;\lambda\rangle$, with 
$\hbar (n-1/2)<S<\hbar(n+1/2)$, ``collapse" to the state $\Psi_n$.

From (5.9) and (5.18$'$) we see that 
when $\lambda \rightarrow 0$, the behavior of the states
$|S,\varphi;\lambda\rangle$ is equivalent to the corresponding behavior of 
the wave functions of E-quantization scheme given by (4.14$''$).

\vspace*{1cm}
\section{Quantum Distribution Functions and a Measurement Procedure}
\setcounter{equation}{0}
\noindent
In this section we consider the physical interpretation of wave
functions $\Psi_{ph}(\xi )$. 
For simplicity we refer again to the equation (5.2)
and assume that the functions $f(\xi )$ and $ g(\xi )$ are 
non-commuting observables 
($\{f, g\}\neq 0$) on the two dimensional phase space $\cal M$. 
We assume as well that solutions of (5.2) 
$\Psi_\epsilon \equiv\Psi_{ph}(\xi )$ define the physical Hilbert space
as the subspace of ${\cal L}_2({\cal M})$. 
To emphasize dependence on 
the observables $f$, $g$ and on the parameter $\epsilon$,
we denote this physical Hilbert space here by ${\cal H}_\epsilon (f, g)$.

On ${\cal H}_\epsilon (f, g)$ the operators $\hat f$ and $\hat g$ have the form
(5.6), where the operator $\hat z^+$ acts on wave functions
$\Psi_{ph} (\xi )$ as the multiplication by
$z^*(\xi )=f(\xi )+i\epsilon g(\xi )$,
and the operator $\hat z$ is it's Hermitian conjugated.
Then, for mean values of these operators we get
\be
 \langle\Psi_{ph}|\hat f|\Psi_{ph}\rangle =
\int d\mu (\xi )~ |\Psi_{ph}(\xi )|^2 f(\xi )~~~~~~
 \langle\Psi_{ph}|\hat g|\Psi_{ph}\rangle =
\int d\mu (\xi )~ |\Psi_{ph}(\xi )|^2 g(\xi )
\ee 
We see, that $|\Psi_{ph}(\xi )|^2$ can be interpreted as some
``distribution function" on the phase space $\cal M$. 

For further investigation we introduce the modulus and phase of 
wave functions $\Psi_{ph}(\xi )$
\be
\Psi_{ph}(\xi )=e^{i\alpha (\xi )}\sqrt{\rho (\xi )}
\ee
From (5.2) and (6.2) we have two real equations \footnote 
{Note, that there are unessential singularities
in the points $\xi_\alpha$, where $\rho (\xi_\alpha )=0$.}
\be
V_f\alpha +\frac{\epsilon}{2}V_g(\log \rho ) =\frac{1}{\hbar}\theta (V_f)
~~~~~~~~~
V_g\alpha -\frac{1}{2\epsilon}V_f(\log \rho ) =\frac{1}{\hbar}\theta (V_g)
\ee 
where $V_f$ and $V_g$ are the corresponding Hamiltonian vector fields
(see (1.1)). 

One can check a validity of the following relations
$$
[V_f, V_g]=V_{\{f,g\}}=
\frac{\{\{f,g\}, g\}}{\{f, g\}}V_f
-\frac{\{\{f,g\}, f\}}{\{f, g\}}V_g
$$
and
$$
V_f\theta (V_g)-V_g\theta (V_f) = \{f, g\} +\theta (V_{\{f, g\}})
$$
Using these relations, we can exclude the function 
$\alpha (\xi )$ from (6.3), and obtain the equation only for $\rho (\xi )$
\be
\left [{\frac{\hbar}{2\epsilon}}{\left (\frac{1}{\{f, g\}}V_f\right )^2 }
+\frac{\hbar\epsilon}{2}{\left (\frac{1}{\{f, g\}}V_g\right )^2}\right ]
\log\rho 
=-\frac{1}{\{f, g\}}
\ee
Note, that in variables $f$, $g$ this equation takes the form
\be
\frac{\hbar}{2}\left (\frac{1}{\epsilon}\partial^2_g +
\epsilon\partial^2_f\right ) \log\rho =-\frac{1}{\{f, g\}}
\ee
where, the Poisson bracket $\{f, g\}$ can be considered
as the function of $f$ and $g$. 

Any solution of (6.4) $\rho (\xi )$  defines corresponding phase 
$\alpha (\xi )$ up to the integration constant (see (6.3)). This constant
phase factor is unessential for physical states (6.2), and respectively
there is one to one correspondence between the ``distribution
functions" $\rho (\xi ) =|\Psi_{ph}(\xi )|^2$ and the pure states
described by a projection operator 
$\hat P_{\Psi_{ph}}\equiv |\Psi_{ph}\rangle\langle\Psi_{ph}|$
\be
\rho (\xi ) \longleftrightarrow \hat P_{\Psi_{ph}}
\ee
With this remark we can use the index $\rho $ for corresponding 
pure states as well: $\hat P_{\Psi_{ph}}\equiv \hat P_\rho$.

From (6.2) and (5.9) we have
\be
\rho (\xi )= |\Psi (\xi )|^2=
\langle\chi_{z(\xi )}|\Psi_{ph}\rangle\langle\Psi_{ph}|
\chi_{z(\xi )}\rangle
\langle\chi_{z(\xi )}|\hat P_\rho |\chi_{z(\xi )}\rangle
\ee
where $|\chi_{z(\xi )}\rangle$ is a coherent state related to the observables
$f$ and $g$ (see (5.8), (5.12)). 
If one introduces the covariant symbol
$P_\rho (\xi )$ of the projection operator
$\hat P_{\rho}$ 
\be
P_\rho (\xi )\equiv
\frac{\langle\chi_{z(\xi )}|\Psi_{ph}\rangle\langle\Psi_{ph}|
\chi_{z(\xi )}\rangle}
 {\langle\chi_{z(\xi )}|\chi_{z(\xi )}\rangle}
\ee
then from (6.7) we have $\rho (\xi )=P_\rho (\xi )||\chi_{z(\xi )}||^2$,
and correspondence (6.6) describes well known 
connection between operators and
their covariant symbols (see [13]). 

For any observable $F(\xi )$ one can introduce the corresponding 
operator $\hat F$ acting on the physical Hilbert space
${\cal H}_\epsilon (f, g)$, and standard quantum mechanical 
mean values are calculated by
\be
\langle \Psi_{ph}|\hat F|\Psi_{ph}\rangle =
Tr(\hat F\hat P_{\rho })\equiv 
\langle \hat F\rangle_\rho 
\ee
Let us introduce a new mean values $\bar F_\rho$:
$$
\bar F_\rho \equiv \int d\mu (\xi )~ F(\xi )\rho (\xi ) 
\eqno (6.9')
$$
The connection between  
mean values $\langle \hat F\rangle_\rho$  and $\bar F_\rho $ 
generally is complicated
and can be done only as an expansion in powers of $\hbar$. But,
for $F=f$ and $F=g$ these mean values are the same for an arbitrary
state $\rho$ (see (6.1))
\be
\bar f_\rho =\langle \hat f\rangle_\rho 
~~~~~~~\bar g_\rho =\langle \hat g\rangle_\rho 
\ee
Using again (5.6), for the operators $\hat f^2$ and $\hat g^2$ 
we obtain
\be
\bar {f^2_\rho} =\langle \hat f^2\rangle_\rho +
\frac{\epsilon\hbar}{2} \langle \hat C\rangle_\rho 
~~~~~~~~~\bar {g^2_\rho} =\langle \hat g^2\rangle_\rho +
\frac{\hbar}{2\epsilon} \langle \hat C\rangle_\rho 
\ee
where the operator $\hat C$ is the commutator
\be
\hat C =\frac{i}{\hbar}[\hat f, \hat g]
\ee
(see (5.14)). 

Quadratic fluctuations calculated for the mean values 
(6.9) and (6.9$'$) respectively are
\be
(\Delta \hat F)^2_\rho =
\langle \hat F^2\rangle_\rho -
\langle \hat F\rangle_\rho^2
\ee
and
$$
(\Delta F)^2_\rho =
 \bar {F^2_\rho} 
- (\bar {F}_\rho )^2 
\eqno (6.13')
$$ 
Then, from (6.10) and (6.11) we have
\be
(\Delta f)^2 = 
(\Delta \hat f)^2 +\frac{\epsilon\hbar}{2}\langle\hat C\rangle
~~~~~~  
(\Delta g)^2 = 
(\Delta \hat g)^2 +\frac{\hbar}{2\epsilon}\langle\hat C\rangle
\ee

In general, a quantum system is not in a pure state
and it is described by a density matrix operator  
$\hat \rho$ [14] which is Hermitian and
semi-positive
($\langle \psi|\hat\rho |\psi\rangle \geq 0$, for  
any state $|\psi\rangle$), and it has the unit trace.
Respectively, any density matrix operator has the spectral expansion
\be
\hat\rho =\sum_n c_n |\psi_n\rangle\langle\psi_n|
\ee
where $|\psi_n\rangle$ are the ortho-normal eigenvectors of $\hat\rho$,
$c_n$ are the corresponding positive ($c_n>0$) eigenvalues, and
$\sum_n c_n =1$.

Similarly to (6.7) we can introduce
the ``distribution function" $\rho (\xi )$
connected with the covariant symbol of $\hat\rho$
\be
\rho (\xi )\equiv 
\langle\chi_{z(\xi )}|\hat\rho|\chi_{z(\xi )}\rangle
\ee
Using the spectral expansion (6.15) we get that a ``distribution function"
of a mixed state  can be expressed as a convex combination
of ``distribution functions" of pure states
\be
\rho (\xi )=\sum_n c_n \rho_n (\xi )~~~~~~~~~(0<c_n<1)
\ee
One can easily check that the relations (6.10), (6.11) and (6.14)
are valid for the mixed states as well.

From (6.16)-(6.17) we see that in general, a ``distribution function" 
$\rho (\xi )$ is a smooth, non-negative function on the phase space $\cal M$,
and it satisfies the standard condition of 
classical distributions
\be
\int d\mu (\xi )~ \rho (\xi ) =1
\ee
Note, that for the pure states the class of functions 
$\rho (\xi )$ essentially depend on 
the value of the parameter $\epsilon$ and on the choice
of observables $f$ and $g$. 
Indeed, for pure states this class is defined by 
solutions of equation (6.4), where this dependence is apparent.
Therefore, sometimes it is convenient to indicate this dependence explicitly:
$\rho (\xi )\equiv \rho (\xi |f,g;\epsilon )$.
In the limit $\epsilon \rightarrow 0$
``distribution functions" $\rho (\xi |f,g;\epsilon )$ become 
singular (see (6.4)), and if the corresponding
operator $\hat f$ has the discrete spectrum $f_n$, then in this limit 
functions $\rho (\xi |f,g;\epsilon)$ should collapse to the
points of this spectrum $f_n$ (see the end of Sections 4 and 5
and the remark below).

Thus, for a given $f(\xi )$, $g(\xi )$ and $\epsilon$ we have 
``distribution functions" $\rho (\xi |f,g;\epsilon)$ 
which look like classical ones,
and at the same time they describe all possible quantum states
uniquely. These functions form a convex set, and corresponding
boundary points satisfy equation (6.4).
Such functions $\rho (\xi |f,g;\epsilon)$ 
we call the quantum distribution functions.

We can compare a function $\rho (\xi |f,g;\epsilon)$ 
to a Wigner function $\rho_w(\xi )$,
which is the Weil symbol of a density matrix operator [16].
For any Wigner function $\rho_w(\xi )$ we have the ``classical" formula
for quantum mechanical mean values
\be
\langle \hat F\rangle_\rho =
\int d\mu (\xi )~ F(\xi )\rho_w (\xi ) 
\ee
Though this formula is valid for an arbitrary observable $F(\xi )$,
nevertheless  Wigner functions can not be interpreted
as a function of probability density.
\footnote {Due to uncertainty principle
there is no such function on the phase space of a quantum system.}
 In general, it is even negative
in some domain of a phase space. It should be noted also, 
that a Wigner function
is defined only for a ``flat" phase space (${\cal M}={\cal R}^{2N}$) and 
cartesian coordinates.

A quantum distribution function 
$\rho (\xi |f,g;\epsilon)$ can be considered for almost arbitrary $f$,
$g$ ``coordinates". It is always positive, but the
``classical" formula (6.19) (with substitution $\rho_w$ by $\rho$)
is valid only for the functions $F=f$ and $F=g$.

For a physical interpretation of quantum distribution functions
$\rho (\xi |f,g;\epsilon)$ we consider again Example 1
(see Section 3) with ${\cal M}= {\cal R}^2$, $f\equiv q$, $g\equiv -p$.

In this case (6.14) takes the form
\be
(\Delta q)^2 = 
(\Delta \hat q)^2 +\frac{\epsilon\hbar}{2}
~~~~~~  
(\Delta p)^2 = 
(\Delta \hat p)^2 +\frac{\hbar}{2\epsilon}
\ee
where $(\Delta \hat q)^2$ and $(\Delta \hat p)^2$ 
are usual quantum mechanical
quadratic fluctuations of coordinate and momentum. Since our quantum theory
for any $\epsilon >0 $ is unitary equivalent to the coordinate (and momentum)
representation, the quadratic fluctuations 
$(\Delta \hat q)^2$ and $(\Delta \hat p)^2$  can be calculated 
also by
\bea
(\Delta \hat q)^2 = 
\int dq~ q^2|\psi (q)|^2 -
\left (\int dq~ q|\psi (q)|^2\right )^2 \nn \\
(\Delta \hat p)^2 = 
\int dp~ p^2|\tilde \psi (p)|^2 -
\left (\int dp~ p|\tilde\psi (p)|^2\right )^2 
\eea
where $\psi (q)$ and $\tilde \psi (p)$ are wave functions
of some pure state $\rho $
in the coordinate and in the momentum representations respectively.
The function $\tilde \psi (p)$ is the Fourier transformation of $\psi (q)$,
and it is well known, that  fluctuations (6.21) satisfy the Heisenberg
uncertainty relation
\be
(\Delta \hat p)^2 (\Delta \hat q)^2 \geq \frac{\hbar^2}{4}
\ee
From (6.20) we have
\be
(\Delta q)^2 \geq \frac{\epsilon\hbar}{2},~~~~~~
(\Delta p)^2 \geq \frac{\hbar}{2\epsilon}~~~~~~
\ee
and using (6.22) we also get
\be
(\Delta p)^2 (\Delta q)^2 \geq {\hbar^2}
\ee

Let us introduce the functions
\be
\rho_\epsilon (q) \equiv \int \frac{dp}{2\pi\hbar}\rho_\epsilon (p,q)
~~~~~~~~
\tilde\rho_\epsilon (p) \equiv \int \frac{dq}{2\pi\hbar}\rho_\epsilon (p,q)
\ee
where $\rho_\epsilon (p,q)$ is a quantum distribution function of this 
example. For pure states\footnote {Generalization to mixed states is 
straightforward.} $\rho_\epsilon (p,q)$ has the form (see (3.12))
\be
\rho_\epsilon (p,q) = \langle p,q;\epsilon|\Psi_{ph}\rangle\langle\Psi{ph}|
p,q;\epsilon\rangle
\ee
and using (3.13$'$) we obtain
\be
\rho_\epsilon (q) = \left (\frac{1}{\pi\hbar\epsilon}\right )^{1/2}
\int dq'~ \exp {\left (-\frac{(q-q')^2}{\hbar\epsilon}\right )}
|\psi(q')|^2
\ee
$$
\tilde\rho_\epsilon (p) = \left (\frac{\epsilon}{\pi\hbar}\right )^{1/2}
\int dp'~ \exp {\left (-\frac{\epsilon(p-p')^2}{\hbar}\right )}
|\tilde\psi(p')|^2
\eqno (6.27')
$$
In the limit 
$\epsilon \rightarrow 0$ and $\epsilon \rightarrow \infty$ we get
\be
{\mbox {when}}~~ \epsilon \rightarrow 0:~~~~~
\rho_\epsilon (q)\rightarrow |\psi (q)|^2,~~~~~
\tilde\rho_\epsilon (p)\rightarrow 0
\ee
$$
{\mbox {when}}~~\epsilon \rightarrow \infty :
 ~~~~~\rho_\epsilon (q)\rightarrow 0,~~~~~
\tilde\rho_\epsilon (p)\rightarrow |\tilde\psi (p)|^2
\eqno (6.28')
$$

From  definitions (6.21) and (6.25) we have the following
correspondence between distribution functions and quadratic fluctuations
\be
|\psi (q)|^2 \leftrightarrow (\Delta\hat q)^2~~~~~~~~~
|\tilde\psi (p)|^2 \leftrightarrow (\Delta\hat p)^2
\ee
$$
\rho_\epsilon (q) \leftrightarrow (\Delta q)^2~~~~~~~~~
\tilde\rho_\epsilon (p) \leftrightarrow (\Delta p)^2~~~~~~~~~
\eqno (6.29')
$$

The function $|\psi (q)|^2$ is a probability density of coordinate
distribution and, in principle, it can be measured. 
The corresponding experiment we denote by $E_q$. Theoretically it is assumed
that in the experiment $E_q$ the coordinate can be measured 
with the absolute precision,  
and a quantum system can be prepared in a given state as many 
times as it is necessary for a good approximation
of the function $|\psi (q)|^2$. 
A statistical distribution of the coordinate, obtained in such experiment, 
is the intrinsic property of a quantum system in a given state:
in general, in a pure state a definite value has some other observable
(for example energy), but not the coordinate.

Similarly, for the momentum distribution function
$|\tilde\psi (p)|^2$ 
we need the experiment $E_p$ with a precise measurement of the momentum.

Thus, in the experiment $E_q$ we can measure the distribution
$|\psi (q)|^2$ and the corresponding quadratic fluctuation 
$(\Delta \hat q)^2$:
$$
E_q \longrightarrow |\psi (q)|^2\longrightarrow (\Delta \hat q)^2
$$
and from the experiment $E_p$ we get 
$|\tilde\psi (p)|^2$
and $(\Delta \hat p)^2$:
$$
E_p \longrightarrow |\tilde \psi (p)|^2\longrightarrow (\Delta \hat p)^2
$$

One possible method for a measurement of a coordinate and a momentum
of a quantum particle is a scattering of a light
on this particle (see [14]).
It is well known, that in such experiment the precise measurement
of the coordinate can be achieved by photons with a very short wavelength 
$\lambda $ (high energy).
On the contrary, for the momentum measurement photons of low energy
are needed.
Theoretically, the experiment $E_q$ is the measurements with photons
of ``zero wavelength": $\lambda \rightarrow 0$,  
and the experiment $E_p$ requires photons of ``zero energy": 
$ \lambda\rightarrow \infty$.
So $E_q$ and $E_p$ are two essentially different experiments.
It should be noted, that in the experiment $E_q$ we measure only the 
coordinate  and 
we have no information about the momentum of a particle. Respectively,
we have not any  momentum distribution function for this experiment.
Similarly, for the absolute precise measurements of the momentum, a particle
can be in any point of the configuration space with equal to each 
other probabilities, and since
the space is infinite, the coordinate distribution function vanishes.

But real experiments, of course, are with photons of finite and 
non-zero wavelength $\lambda$. Experiment with some fixed wavelength
$\lambda$ we denote by $E_\lambda$. In this experiment there is the error 
$\Delta_q$ in measuring of the coordinate and this error is proportional
to the wavelength $\lambda$
(see [14])
\be
\Delta_q=\alpha \lambda
\ee
Here $\alpha $ is a dimensionalless parameter of order 1 ($\alpha \sim 1$).

The  momentum of a photon with a wavelength $\lambda$ is 
$$p_\lambda =\frac{2\pi\hbar}{\lambda}$$ and the error of momentum 
measurement is proportional to this momentum 
\be
\Delta_p=\beta\frac{\hbar}{\lambda}
\ee
where $\beta$ is the parameter similar to $\alpha$ ($\beta \sim 1$).

Then, for the total quadratic fluctuations we can write
\bea
(\Delta_tq)^2 =(\Delta\hat q)^2+(\Delta_q)^2=
(\Delta\hat q)^2+\alpha^2\lambda^2 \nn \\
(\Delta_tp)^2 =(\Delta\hat p)^2+(\Delta_p)^2=(\Delta\hat p)^2+
\frac{\beta^2\hbar^2}{\lambda^2}
\eea
Thus, in the experiment $E_\lambda$ we have two kind of fluctuations:
the first one 
($(\Delta\hat q), (\Delta\hat p)$)
is the intrinsic property of a quantum system, and the second
($(\Delta_q), (\Delta_p)$) is related to the measurement procedure.
As it is well known, the fluctuations $(\Delta\hat q)$ and $(\Delta\hat p)$ 
satisfy Heisenberg uncertainty principle (6.22).
Assuming that for the ideal experiment $\alpha =\beta =1/{\sqrt {2}}$, 
we can fix  the uncertainties of a measurement procedure by
\be
\Delta_q\Delta_p =\frac{\hbar}{2}~~~~~~~
\frac{\Delta_q}{\Delta_p}=\frac{\lambda^2}{\hbar}
\ee
With this assumption, from (6.20) and (6.32), we can write
\be
(\Delta_tq)^2 =(\Delta q)^2~~~~~~(\Delta_tp)^2=(\Delta p)^2
\ee
and the parameter $\epsilon$  and the wavelength $\lambda$ are related
by
\be
\lambda =\sqrt {\epsilon\hbar}
\ee
Recall that the quadratic fluctuations
$(\Delta q)^2$ and $(\Delta p)^2$ 
are calculated by the mean values of the function
$\rho_\epsilon (p,q)$ (see (6.20)).
Taking into account (6.32)-(6.35) one can suppose that these fluctuations
respectively are the total quadratic fluctuations 
of the coordinate and the momentum measured in the experiment $E_\lambda$.
As it was mentioned, in this experiment we have some unavoidable
non-zero measurement error both for the coordinate
and the momentum, and the parameter $\epsilon $ fixes the ratio of these errors
(see (6.33), (6.35))
$$
\epsilon =\frac{\Delta_q}{\Delta_p}
\eqno (6.35')
$$
It is worth noting that in the experiments $E_\lambda$ 
one can carry out the separate measurement of
coordinate and momentum  as well as do it
simultaneously.
Then, the function
$\rho_\epsilon (q)$ (see (6.27) and (6.29$'$))
can be interpreted, as a distribution
of the coordinate obtained in the experiment $E_\lambda$. 
Similarly, the function $\tilde\rho_\epsilon (p)$
corresponds to the momentum measurements in $E_\lambda$.
In the limit 
$\lambda \rightarrow 0$ ($\epsilon \rightarrow 0$ )
we get the experiment $E_q$ with the coordinate distribution 
$\rho_0(q)=|\psi (q)|^2$ only, and in the opposite limit 
$\lambda \rightarrow \infty$ 
(the experiment $E_p$) 
only the distribution $|\tilde\psi (p)|^2$ remains (see (6.28))\footnote
{The distributions $\rho_\infty (q)$ and $\tilde\rho_0 (p)$ are 
degenerated to zero functions.}

Now, it is natural to suppose that the quantum distribution function
$\rho_\epsilon (p,q) =|\Psi_\epsilon (p,q)|^2$ is the distribution 
obtained in the experiment $E_\lambda$ with simultaneous 
measurements of the coordinate and the momentum.

This idea can be easily generalized 
assuming that the quantum distribution function $\rho(\xi |f,g:\epsilon)$
is the distribution on the phase space obtained in some ideal experiment
with simultaneous measuring of $f$ and $g$ observables. 
In such experiment we have the unavoidable errors  
$\Delta_f$ and $\Delta_g$
connected with the measurement procedure with micro-objects.
For corresponding fluctuations there is the additional uncertainty 
principle (see (6.33)), and the parameter $\epsilon$ specifies the experiment
by fixing the ratio of the errors $\epsilon =
\Delta_f/\Delta_g$.

If the function $\rho (\xi )\equiv\rho (\xi |f,g;\epsilon)$ is really 
measurable, then in the limit $\epsilon \rightarrow 0$ this function
$\rho (\xi )$ should describe the experimental distribution of the
exact measurement of the observable $f$. It is obvious that for 
the observable $f$ with 
discrete spectrum corresponding function $\rho (\xi )$ should be localized
in the points of this spectrum. Thus, by asymptotics of quantum
distribution functions one can obtain the spectrum of the physical
observables (see (4.14$''$)).

We see that quantum distribution functions 
can play  some fundamental role for the interpretation
of quantum theory. It is natural to try to formulate  quantum 
mechanics in terms of these distribution functions, especially as, 
they describe all possible states of a quantum system uniquely. 
But for this it is worthwhile to have an independent 
(without referring to the Hilbert space) 
description of the set of functions 
$\rho (\xi )\equiv\rho (\xi |f,g;\epsilon )$. 
Corresponding functions
are positive, satisfying (6.18), and at the same time they essentially
depend on the choice of observables $f$ and $g$ and of the parameter $\epsilon$.
On the other hand, the set of physical states 
is a convex one, were the boundary points are 
the pure states. So for the description of our set we need to specify the 
distribution functions of pure states, but the latter
are given as the solutions of (6.4).
Thus, in this approach 
the important role plays the equation (6.4).
Actually it describes the set of all physical states and, respectively,
it contains the information about quantum uncertainties 
both the intrinsic and the experimental ones.

Note, that on the left hand side of the corresponding equation 
there is the Laplace operator (see (6.4)-(6.5))
and we have some induced metric structure on the phase space $\cal M$. 
It is remarkable, that this metric structure 
is related to the experimental errors. 
Indeed, in case of Example 1 these
errors are (see (6.33-(6.35))
$$
\Delta_q = \sqrt {\frac{\epsilon\hbar}{2}} ~~~~~~~~~~
\Delta_p = \sqrt {\frac{\hbar}{2\epsilon}} 
$$
and it is easy to see that corresponding equation (6.5) takes the form
\be
\left (\Delta_q^2 \partial^2_q + 
\Delta_p^2 \partial^2_p\right )\log\rho = -1 
\ee
Such kind of phase space ``shadow" metric was introduced in [4c].

If the equation for the quantum distribution functions of pure states
has really the fundamental character, then one might expect that 
it can be derived from some general principle. 
A suitable principle could be the minimization of
certain functional,
and we arrive to the problem of construction of the corresponding 
functional.
 Since the minimization should be achieved on pure states,
it is natural to interpret such functional as the entropy of a
quantum system. Respectively, one candidate for such functional is the standard
quantum mechanical
entropy $S=-Tr(\hat \rho\log\hat\rho)$ which can be expressed as the functional
of $\rho (\xi )$.

It seems, that this and other above mentioned problems are interesting and
need further investigation.

\vspace*{1cm}
\begin{center}{\Large \bf Appendix A}
\end{center}
\noindent
Let $f, g$ be two non-commuting observables and $\hat {\Phi}_f, \hat {\Phi}_g$ 
the corresponding constraint operators (2.7). 
As it was mentioned, these operators are Hermitian on the Hilbert space 
${\cal H}\equiv {\cal L}_2({\cal M})$. 
Suppose, that the equation (see (2.9))
$$
(\hat {\Phi}_f +i\epsilon\hat {\Phi}_g )|\Psi_\epsilon\rangle=0
\eqno (A.1)
$$
has normalizable solutions for any $\epsilon \in (0,\delta)$, where
$\delta$ is some positive number. The solutions with fixed $\epsilon $
form some subspace ${\cal H}_\epsilon$ of the Hilbert space $\cal H$.
We assume, that each subspace can be represented as 
${\cal H}_\epsilon =F_\epsilon {\cal H}_0$, where 
${\cal H}_0$ is some linear space, and $F_\epsilon$ is a linear
invertible map
$$
F_\epsilon : {\cal H}_0 \rightarrow {\cal H}_\epsilon ~~~~~~~~~~
F^{-1}_\epsilon : {\cal H}_\epsilon \rightarrow {\cal H}_0
\eqno (A.2)
$$
In practical applications the linear space
${\cal H}_0$
automatically arises from the form of the general solution of (A.1);
only it should be specified from the condition of square integrability
of corresponding functions $\Psi_\epsilon =F_\epsilon\psi$, 
where $\psi\in{\cal H}_0$.
For example, in case of eq. (3.3), 
the general solution (3.4)-(3.5) is described 
by the space of polynomials $P(\xi )$, and it can be
interpreted as ${\cal H}_0$.
The representation (3.12) and (3.15) of the same solutions is different,
and in that case, the space ${\cal H}_0$ obviously is ${\cal L}_2({\cal R}^1)$.
As for the general solution (3.25)-(3.26), 
the space ${\cal H}_0$ is a space of Fourier modes $c_n,~ n\in Z$, with
$\sum |c_n|^2 < \infty$ (see (3.27)).

The space of linear functionals on the Hilbert space
${\cal H}$ is called the dual (to ${\cal H}$) space, and we denote it
by ${\cal H}^*$. From our definitions we have 
$$
\Psi_\epsilon =F_\epsilon\psi\in{\cal H}_\epsilon \subset
{\cal H}\subset{\cal H}^* 
$$
Suppose that the set of vectors $F_\epsilon\psi$
with any fixed $\psi\in {\cal H}_0$, has the limit 
($\epsilon\rightarrow 0$) in the dual space ${\cal H}^*$, 
and this limit defines the vector $\psi_*\in{\cal H}^*$
$$\lim \limits_{\epsilon \to 0}
F_\epsilon \psi =\psi_*
\eqno (A.3)
$$
Such linear functional $\psi_*$ usually is unbounded, and
the limit in (A.3) means that for any $\Psi\in
{\cal H}$ we have\footnote {As an unbounded functional $\psi_*$ is not
defined for an arbitrary $\Psi\in{\cal H}$, but the 
domain of definition of $\psi_*$ should be
everywhere dense set in $\cal H$.}
$$
\lim \limits_{\epsilon \to 0}
\langle F_\epsilon \psi|\Psi\rangle =
\psi_*(\Psi )
\eqno (A.3')
$$
where $\psi_*(\Psi )$ denotes the value of the functional $\psi^*$
on the corresponding vector $\Psi\in{\cal H}$. 
If we change the map $F_\epsilon$ by
$$
F_\epsilon\rightarrow\tilde F_\epsilon =a(\epsilon )F_\epsilon
$$
where $a(\epsilon )$ is some ``scalar" function of the parameter
$\epsilon $, then the new map $\tilde F_\epsilon$ provides representation
of the subspace ${\cal H}_\epsilon$ in the same form:
${\cal H}_\epsilon =\tilde F_\epsilon {\cal H}_0$. It is obvious
that existence of the limit in (A.3) essentially depends on the
suitable choice of the normalizable function $a(\epsilon )$.
 
The action of some operator $\hat O$ on the functional $\psi_*$ can be defined
by 
$$
\hat O\psi_*(\Psi )\equiv \psi_*(\hat O^+\Psi )
\eqno (A.4)
$$
where $\hat O^+$ is the Hermitian conjugated to $\hat O$. 

The norm $||\Psi_\epsilon ||$ of the vectors $\Psi_\epsilon =F_\epsilon\psi$,
with fixed $\psi$, usually diverges when 
$\epsilon \rightarrow 0$, but if we assume that 
$$
\epsilon||F_\epsilon\psi || \rightarrow 0 
\eqno (A.5)
$$
then we can prove that $\psi_*$ satisfies the equation $\hat \Phi_f~\psi_*=0$.
Indeed, from (A.3)-(A.5) we have
$$
\hat \Phi_f\psi_*(\Psi )=\psi_*(\hat \Phi_f\Psi )
=\lim \limits_{\epsilon \to 0}\langle F_\epsilon \psi|\hat \Phi_f\Psi\rangle
=\lim \limits_{\epsilon \to 0}\langle \Psi_\epsilon|\hat \Phi_f\Psi\rangle 
=\lim \limits_{\epsilon \to 0}i\epsilon \langle \Psi_\epsilon|\hat \Phi_g\Psi\rangle =
0
\eqno (A.6)
$$
where we take into account that the function $\Psi_\epsilon =F_\epsilon\psi$
satisfies  (A.1).
Thus, (A.3) defines the map $F_*:{\cal H}_0\rightarrow{\cal H}^*$,
and corresponding functionals $\psi_*=F_*\psi$ satisfy condition (2.8). 

Further, let us assume, that $F_*\psi \neq 0$, whenever $\psi \neq 0$. 
Then, the space ${\cal H}_{ph}\equiv F_*{\cal H}_0$, 
as the linear space, will be isomorphic to ${\cal H}_0$, 
and, respectively, isomorphic to each ${\cal H}_\epsilon$ as well (see (A.2)).

If for $\forall~ \epsilon_1, \epsilon_2 \in (0,\delta )$ the map 
$$ F_{\epsilon_2} F^{-1}_{\epsilon_1} : {\cal H}_{\epsilon_1} \rightarrow 
{\cal H}_{\epsilon_2}
\eqno (A.7)
$$
is an unitary transformation, then one can introduce the Hilbert 
structure on ${\cal H}_0$ and ${\cal H}_{ph}$
by definition of the scalar product
$$
\langle \psi_2|\psi_1\rangle \equiv
\langle F_*\psi_2|F_*\psi_1\rangle \equiv
\langle F_\epsilon \psi_2|F_\epsilon \psi_1\rangle 
\eqno (A.8)
$$
It is obvious that in case of unitarity of transformations (A.7)
the scalar product (A.8) is independent on the choice of the parameter 
$\epsilon$,
and the corresponding Hilbert structure is a natural.
But, in general, transformation (A.7) is not the unitary one, and there is
no some special Hilbert structure on
${\cal H}_0$. Respectively, we have the problem 
for the scalar product on the space ${\cal H}_{ph}$, especially as, 
corresponding functionals are unbounded and have the ``infinite norm"
in the Hilbert space $\cal H$. 

Note, that for the general solutions (3.4)-(3.5) corresponding transformation
(A.7) is not the unitary one, while the general solution (3.11)-(3.12), (3.15)
provides unitarity explicitly
$$
\tilde \Psi_{\epsilon_2} (p,q)= 
\int \frac{dpdq}{2\pi\hbar}~\langle p,q;\epsilon_2| p',q';\epsilon_1\rangle 
\tilde \Psi_{\epsilon_1}(p',q')
$$

Now, we describe some procedure
for the solution of scalar product problem in that general case too.

In ordinary quantum mechanics a physical state is represented by 
a ray in a Hilbert space,
and all vectors on the same ray are physically indistinguishable.
So, if we suppose that the vector $|\psi_*\rangle$ has some norm
$||\psi_*||$, then the normalized vector
$$
|\psi_*\rangle\rangle\equiv\frac{|\psi_*\rangle}
{||\psi_*||}
\eqno (A.9)
$$ 
describes the same physical state. It is just the scalar 
product of such normalized vectors that has the physical meaning.
 Up to the phase factor,
this scalar product describes the
``angle" between the rays, and defines the probability amplitude.

We introduce the scalar product of such normalized vectors by
$$
\langle\langle\psi_{2*} |\psi_{1*}\rangle\rangle \equiv
\lim \limits_{\epsilon \to 0} \frac 
{\langle\Psi_{2\epsilon}|\Psi_{1\epsilon}\rangle }
{||\Psi_{2\epsilon}||~||\Psi_{1\epsilon}||}
\eqno (A.10)
$$
where the limits of $|\Psi_{1\epsilon}\rangle$ and $|\Psi_{2\epsilon}\rangle$ 
respectively are the functionals $|\psi_{1*}\rangle $ and $|\psi_{2*}\rangle $
(see (A.3)), and the latter are related to 
$|\psi_{1*}\rangle\rangle$ and $|\psi_{2*}\rangle\rangle$ by (A.9).
When the limit (A.10) exists, it should define the scalar
product of the normalized physical states. 
Then, the scalar product for arbitrary vectors can be obtained 
uniquely up to a rescaling. 

It is obvious, that in case of unitarity of transformations (A.7), 
the definitions of scalar product (A.8) and (A.10) are equivalent.

Note, that the described scheme for the definition of scalar product
of physical states (2.8) can be generalized for other constrained systems
as well.

\vspace*{1cm}
\begin{center}{\Large \bf Appendix B}
\end{center}
\noindent
Let us consider a symplectic manifold ${\cal M}$ with
global coordinates $\xi^k,~ (k=1,...,2N)$ and constant
symplectic matrix: 
$\partial_j\omega^{kl}=0$, where 
 $\omega^{kl}= -\{\xi^k ,\xi^l \}$ (see (1.2$'$)). The simple example 
of such $\cal M$ is ${\cal R}^{2N}$ with canonical coordinates.

For the global coordinates $\xi^k$ we can introduce the corresponding
constraint functions $\Phi_{\xi^k}$, and from (1.6$'$)-(1.7)
we get
$$
\Phi_{\xi^k} =\Phi^k =\omega^{kl} (P_l - \theta_l)
\eqno {(B.1)}
$$
Then, (1.8) takes the form
$$
\{ \Phi^k,\Phi^l\}_* =\omega^{kl} 
~~~~~~~~~\{f,\Phi^k\}_* = -\omega^{kl}\partial_lf
\eqno {(B.2)}
$$
where $f(\xi )$ is any observable on ${\cal M}$, but in (B.2) it
is considered as a function on $T^*{\cal M}$ with natural
extension (see remarks after eq. (1.8)). 

Let us add to the function $f(\xi )$ the term linear in constraints
$\Phi^k$
$$
f(\xi )\rightarrow  f^{(1)} =f(\xi ) +A^{(1)}_l(\xi )\Phi^l
\eqno {(B.3)}
$$
and choose the functions $A^{(1)}_l(\xi )$ to satisfy the condition
$$
\{f^{(1)},\Phi^k\}_* = 
B^{(1)k}_l(\xi )
\Phi^l
\eqno {(B.4)}
$$
This means, that the right hand side of (B.4)
should contain the constraints $\Phi^k$ only in the first degree.
From this condition the functions 
$A^{(1)}_l(\xi )$ and $B^{(1)k}_l(\xi )$ are defined uniquely
$$
A^{(1)}_l(\xi ) =-\partial_l f(\xi )~~~~~~~~
B^{(1)k}_l(\xi )=\omega^{kj}\partial^2_{jl}f(\xi )
\eqno {(B.5)}
$$
It is obvious, that $f^{(1)}=R_f$, and (B.4)-(B.5) are equivalent
to (2.2) and (2.5) with constant symplectic matrix $\omega^{kl}$.
We can continue this ``deformation" procedure
$$
 f^{(1)} \rightarrow  
f^{(2)} = 
f^{(1)} +\frac{1}{2}
A^{(2)}_{lj}(\xi )
\Phi^l\Phi^j
\eqno {(B.6)}
$$
demanding
$$
\{f^{(2)},\Phi^k\}_* = 
B^{(2)k}_{lj}(\xi )
\Phi^l\Phi^j
$$
Then, for the functions $A^{(2)}_{lj}(\xi )$ and $B^{(2)k}_{lj}(\xi )$
we have
$$
A^{(2)}_{lj}(\xi )=\partial^2_{lj}f(\xi )~~~~~~~~
B^{(2)k}_{lj}(\xi )=-\frac{1}{2}\omega^{ki}\partial^3_{ilj}f(\xi )
$$
Generalizing for arbitrary $n$, we get
$$
f(\xi )\rightarrow  f^{(n)} =f(\xi ) +\sum_{a=1}^{n}~ 
\frac{1}{a!}
A^{(a)}_{k_1...k_a}(\xi )\Phi^{k_1}....\Phi^{k_a}
\eqno {(B.7a)}
$$
where
$$ 
A^{(a)}_{k_1...k_a}(\xi ) = (-)^a\partial^{(a)}_{k_1...k_a}f(\xi )
\eqno {(B.7b)}
$$
and
$$
\{f^{(n)},\Phi^k\}_* = 
\frac{(-)^{n+1}}{n!}\omega^{kl}\left (\partial
^{(n+1)}_{lk_1...k_n}f(\xi )\right )\Phi^{k_1}....\Phi^{k_n}
\eqno {(B.7c)}
$$
Using this procedure for any observable $f(\xi )$, one can construct
a new function $\tilde f =\lim f^{(n)}~ (n\rightarrow \infty )$, which
commutes with all constraints 
$\Phi^k~(k=1,...,2N)$, and on the constraint surface 
($\Phi^k=0~(k=1,...,2N)$) it is equal to $f(\xi )$.

A similar procedure can be accomplished on the quantum level as well,
taking into account operators ordering and self-adjoint conditions.
But, when the symplectic matrix $\omega^{kl}$ depends on coordinates
$\xi^k$, the described procedure fails for some observables $f(\xi )$,
even on the classical level.
For the illustration let us consider a simple example on a half plane
with coordinates $(p,q),~ p>0$, and the canonical 1-form $\theta =pdq$. 
If we take the coordinates 
$\xi^1={p^2}/2,~\xi^2=q$ (which are global here), 
then the corresponding constraints
$\Phi^1 = p^2-pP_q, ~ \Phi^2=P_p$
have the commutation relations
$$
\{\Phi^2, \Phi^1\}=p+\frac{1}{p}\Phi^1
\eqno {(B.8)}
$$
The first deformation of the function $f=q$, as usual,
gives $f^{(1)}=R_q=q-P_p$ and we get
$$
\{f^{(1)},\Phi^1\}_* =-\frac{1}{p}\Phi^1~~~~~~~
\{f^{(1)},\Phi^2\}_* =0
\eqno {(B.9)}
$$
Considering the second deformation  (B.6)
$$
f^{(2)} = 
f^{(1)} +\frac{1}{2}\left (A_{11}(\xi )(\Phi^1)^2 +
2A_{12}(\xi )\Phi^1\Phi^2 +A_{22}(\xi )(\Phi^2)^2\right )
$$
and using commutation relations (B.8)-(B.9), we see, that it is impossible
to cancel the linear (in constraints 
$\Phi^1 $  and $\Phi^2$) terms in the Poisson brackets
$\{f^{(2)},\Phi^1\}_*$ and $\{f^{(2)},\Phi^2\}_*$ simultaneously.

\vspace*{1cm}
\begin{center}{\Large \bf Appendix C}
\end{center}
\noindent
At first we consider minimization of the product of quadratic fluctuations
(see (4.6))
$$
U(\Psi )\equiv
\langle \Psi|\hat {\Phi}_f^2|\Psi\rangle 
\langle \Psi|\hat {\Phi}_g^2|\Psi\rangle 
\eqno {(C.1)}
$$
with the vectors $|\Psi\rangle$ of unit norm
$$
\langle\Psi|\Psi\rangle =1
\eqno {(C.2)}
$$ 
For the minimization of the functional $U(\Psi )$ one can use the
variation principle, considering the variation of 
$|\Psi\rangle$
to be independent of
$\langle\Psi|$. Since we have the subsidiary condition (C.2), from the
variation of (C.1) we obtain
$$
b^2\hat {\Phi}_f^2|\Psi\rangle+a^2\hat {\Phi}_g^2|\Psi\rangle=
c|\Psi\rangle
\eqno {(C.3)}
$$
where
$$
a^2= \langle \Psi|\hat {\Phi}_f^2|\Psi\rangle
~~~~~~~~~b^2= \langle \Psi|\hat {\Phi}_g^2|\Psi\rangle
\eqno {(C.4)}
$$
Multiplying by $\langle\Psi |$, we get $c=2a^2b^2$, and 
the equation (C.3) takes the form
$$
\frac{1}{2a^2}\hat {\Phi}_f^2|\Psi\rangle+\frac{1}{2b^2}
\hat {\Phi}_g^2|\Psi\rangle=|\Psi\rangle
\eqno {(C.5)}
$$
Thus, the solutions of (C.5), which satisfy conditions
(C.4) can provide minimization of the functional $U(\Psi )$.
If there are solutions with different values of the parameters
$a$ and $b$, then we have to choose the solutions with minimal
value of the product $a^2b^2$.

Now we consider minimization of the functional $U_1(\Psi )$ (see (4.7))
$$
U_1(\Psi )\equiv\frac{\langle\Psi|\hat {\Phi}^2_f|\Psi\rangle
\langle\Psi|\hat {\Phi}^2_g|\Psi\rangle}
{\langle\Psi|\hat A|\Psi\rangle^2}
\eqno {(C.6)}
$$

For an arbitrary vector $|\Psi\rangle$ and any real parameter $\epsilon$
we have
$$
\langle\Psi|(\hat {\Phi}_f-i\epsilon\hat {\Phi}_g )
(\hat {\Phi}_f +i\epsilon\hat {\Phi}_g )
|\Psi\rangle\geq 0
\eqno {(C.7)}
$$
The left hand side of this inequality is a second ordered polynomial
in $\epsilon $ 
$$
\epsilon^2\langle\Psi|\hat {\Phi}^2_g|\Psi\rangle
-\hbar\epsilon
\langle\Psi|\hat A|\Psi\rangle +
\langle\Psi|\hat {\Phi}^2_f|\Psi\rangle
$$
and respectively we have
$$
\langle\Psi|\hat {\Phi}^2_f|\Psi\rangle
\langle\Psi|\hat {\Phi}^2_g|\Psi\rangle
\geq \frac{\hbar^2}{4}\langle\Psi|\hat {A}|\Psi\rangle^2
\eqno {(C.8)}
$$
Thus, the minimal value of the functional
$U_1(\Psi )$
could be ${\hbar^2}/4$. If for some $\epsilon$ the equation
$$
(\hat {\Phi}_f +i\epsilon\hat {\Phi}_g )
|\Psi\rangle = 0
\eqno {(C.9)}
$$
has normalizable solution  $|\Psi\rangle =
|\Psi_\epsilon\rangle$,
then, for this 
$|\Psi_\epsilon\rangle$
we have an equality
in (C.7) and (C.8). Respectively, this states
$|\Psi_\epsilon\rangle$, provide minimization of the functional
$U_1(\Psi )$.
But, as it was indicated in section 4, sometimes equation (C.9) has no
normalizable solutions for any real $\epsilon$. In that case, one can
consider minimization problem for 
the functional 
$U_1(\Psi )$
by variation principle, as it was done above for the functional
$U(\Psi )$.
Repeating the same procedure, we get the equation
$$
\frac{1}{2a^2}\hat {\Phi}_f^2|\Psi\rangle+\frac{1}{2b^2}
\hat {\Phi}_g^2|\Psi\rangle -\frac 
{\hat A}{A}|\Psi\rangle =0
\eqno (C.10)
$$
where $a,b,A$ are parameters, and the solution $|\Psi \rangle$
should satisfy (C.4) and the additional condition
$\langle\Psi|{\hat A}|\Psi\rangle=A$ as well.

\vspace{1.cm}

{\bf {Acknowledgments}}
\vspace*{0.5cm}

This work has been supported in part by grants INTS, JSPS,
and RFFR.

The author is most grateful to Professor Y.\ Sobouti for kind hospitality at
Zanjan Institute for Advanced Steadies and for the collaboration.

The main part of this work was done during the visit in Trieste at ICTP. 
The author greatly acknowledges the support of the High Energy 
Section and the Associateship office for their support
and kind attention.

The author thanks  G.\ Chechelashvili, E.S.\ Fradkin,
V.\ Georgiev, G.C.\ Ghirardi, E.\ Gozzi, A.\ Karabekov,
L.\ Lusanna, P.\ Maraner for the interest to this work
and helpful discussions.

\end{document}